\documentclass[aps,prd,amsmath
,eqsecnum            
,preprintnumbers
,nofootinbib         
,twocolumn         
]{revtex4}
\usepackage[dvips]{graphicx}
\usepackage{amsmath}
\usepackage{amsfonts}
\usepackage{amssymb}
\usepackage{longtable}   

\allowdisplaybreaks[4]     

\newcommand{\df}{\ {\overset {\rm def} =}\ }
\newcommand{\dr}[2]{\frac {{\rm d} {#1}} {{\rm d} {#2}}}
\newcommand{\pdr}[2]{\frac {\partial {#1}} {\partial {#2}}}
\newcommand{\dril}[2]{{{\rm d} {#1}} / {{\rm d} {#2}}}

\newcommand{\llim}[1] {\ {\underset {#1} {\longrightarrow}}\ }

\begin{document}

\title{Redshift propagation equations in the $\beta' \neq 0$ Szekeres models}

\author{Andrzej Krasi\'nski}
\affiliation{N. Copernicus Astronomical Centre, Polish Academy of Sciences, \\
Bartycka 18, 00 716 Warszawa, Poland} \email{akr@camk.edu.pl}
\author{Krzysztof Bolejko}
\affiliation{Astrophysics Department, University of Oxford, Oxford OX1 3RH, UK}
\email{Krzysztof.Bolejko@astro.ox.ac.uk}

\date { }

\begin{abstract}
The set of differential equations obeyed by the redshift in the general $\beta'
\neq 0$ Szekeres spacetimes is derived. Transversal components of the ray's
momentum have to be taken into account, which leads to a set of 3 coupled
differential equations. It is shown that in a general Szekeres model, and in a
general Lema\^{\i}tre -- Tolman (L--T) model, generic light rays do not have
repeatable paths (RLPs): two rays sent from the same source at different times
to the same observer pass through different sequences of intermediate matter
particles. The only spacetimes in the Szekeres class in which {\em all} rays are
RLPs are the Friedmann models. Among the proper Szekeres models, RLPs exist only
in the axially symmetric subcases, and in each one the RLPs are the null
geodesics that intersect each $t =$ constant space on the symmetry axis. In the
special models with a 3-dimensional symmetry group (L--T among them), the only
RLPs are radial geodesics. This shows that RLPs are very special and in the real
Universe should not exist. We present several numerical examples which suggest
that the rate of change of positions of objects in the sky, for the studied
configuration,  is $10^{-6} - 10^{-7}$ arc sec per year. With the current
accuracy of direction measurement, this drift would become observable after
approx. 10 years of monitoring. More precise future observations will be able,
in principle, to detect this effect, but there are basic problems with
determining the reference direction that does not change.
\end{abstract}

\maketitle

\section{The motivation}

The quasi-spherical Szekeres solutions have recently begun to be taken seriously
as cosmological models \cite{Bole2006} -- \cite{NTI2010}. For this application,
one has to know the equations obeyed by the redshift. The corresponding equation
for radial null geodesics in the Lema\^{\i}tre -- Tolman (L--T) model
\cite{Lema1933, Tolm1934} was derived long ago by Bondi \cite{Bond1947}, see
also Ref. \cite{PlKr2006}. The generalisation to the Szekeres geometry is
nontrivial because in general there are no radial geodesics in the latter
\cite{BKHC2009, NoDe2007}. Consequently, the transversal components of the ray's
momentum necessarily have to be taken into account, and a set of 3 coupled
differential equations is obtained. These equations can then be applied to
nonradial geodesics in the L--T model.

The purpose of this paper is to derive the redshift propagation equations in a
general\footnote{{\em General} means not only quasi-spherical. The
generalisation to cover the quasi-plane and quasi-hyperbolic cases is immediate,
so it would not make sense to leave it out.} Szekeres model of the $\beta' \neq
0$ family \cite{PlKr2006}, so that they can be numerically solved and applied in
various situations.

In Sec. \ref{Szek}, the Szekeres models are introduced. In Sec. \ref{Bondi} it
is pointed out that the Bondi redshift equation for radial null geodesics in the
L--T model is in fact an approximation, the small parameter being the period of
the electromagnetic wave. The same is true for the equations derived here. In
Sec. \ref{nullgeo}, the general equations of null geodesics in Szekeres models
are presented. In Sec. \ref{Szredshift}, the set of redshift equations for the
Szekeres models is derived. In Sec. \ref{repeat}, conditions are discussed under
which light rays between a given source and a given observer proceed through
always the same intermediate matter particles; such rays are termed ``repeatable
light paths'', RLPs. In Sec. \ref{LTredshift}, the equations of Secs.
\ref{Szredshift} and \ref{repeat} are applied to general null geodesics in the
L--T model and in the associated plane- and hyperbolically symmetric models. It
is shown there that in these models the only RLPs are the radial null geodesics.
Sec. \ref{summary} is a brief summary of the results.

\section{The Szekeres solutions}\label{Szek}

\setcounter{equation}{0}

The Szekeres solutions \cite{Szek1975a,Szek1975b} follow when the metric
\begin{equation}\label{2.1}
{\rm d} s^2 = {\rm d} t^2 - {\rm e}^{2 \alpha(t,r,x,y)} {\rm d} r^2- {\rm e}^{2
\beta(t,r,x,y)} \left({\rm d} x^2 + {\rm d} y^2\right),
\end{equation}
is substituted in the Einstein equations with a dust source, assuming that the
coordinates of (\ref{2.1}) are comoving, so that the velocity field is $u^{\mu}
= {\delta^{\mu}}_0$ (with $(x^0, x^1, x^2, x^3) = (t, r, x, y))$.

There are two families of Szekeres solutions, depending on whether $\beta,_r =
0$ or $\beta,_r \neq 0$. The first family is a simultaneous generalisation of
the Friedmann and Kantowski -- Sachs \cite{KaSa1966} models. Since so far it has
found no useful application in astrophysical cosmology, we shall not discuss it
here (see Ref. \cite{PlKr2006}). After the Einstein equations are solved, the
metric functions in the second family become
\begin{eqnarray}\label{2.2}
{\rm e}^{\beta} &=& \Phi(t, r) {\rm e}^{\nu(r, x, y)}, \nonumber \\
{\rm e}^{\alpha} &=& h(r) \Phi(t, r) \beta,_r \equiv h(r) \left(\Phi,_r + \Phi
\nu,_r\right), \\
{\rm e}^{- \nu} &=& A(r)\left(x^2 + y^2\right) + 2B_1(r) x + 2B_2 (r)y + C(r),
\nonumber
\end{eqnarray}
where the function $\Phi(t, r)$ is a solution of the equation
\begin{equation}\label{2.3}
{\Phi,_t}^2 = - k(r) + \frac {2 M(r)} {\Phi} + \frac 1 3 \Lambda \Phi^2;
\end{equation}
while $h(r)$, $k(r)$, $M(r)$, $A(r)$, $B_1(r)$, $B_2(r)$ and $C(r)$ are
arbitrary functions obeying
\begin{equation}\label{2.4}
g(r) \df 4 \left(AC - {B_1}^2 - {B_2}^2\right) = 1/h^2(r) + k(r).
\end{equation}
The mass density in energy units is
\begin{equation}\label{2.5}
\kappa \rho = \frac {\left(2M{\rm e}^{3\nu}\right),_r} {{\rm e}^{2\beta}
\left({\rm e}^{\beta}\right),_r}; \qquad \kappa = 8 \pi G / c^4.
\end{equation}
Whenever $\left({\rm e}^{\beta}\right),_r = 0$ and $\left(2M{\rm
e}^{3\nu}\right),_r \ne 0$, a shell crossing singularity occurs. It is similar
to the shell crossing singularity in the L--T models, but with a difference. In
a quasi-spherical model a shell crossing may occur along a circle, or, in
exceptional cases, at a single point, and not at a whole surface of constant $t$
and $r$, as was the case in the L--T models.

As in the L--T model, the bang time function follows from (\ref{2.3}):
\begin{equation}\label{2.6}
\int\limits_0^{\Phi}\frac{{\rm d} \widetilde{\Phi}}{\sqrt{- k + 2M /
\widetilde{\Phi} + \frac 1 3 \Lambda \widetilde{\Phi}^2}} = t - t_B(r),
\end{equation}
The solutions of the above equation for $\Lambda \neq 0$ involve elliptic
functions and were first studied by Barrow and Stein-Schabes \cite{BaSS1984}.

As seen from (\ref{2.1}) and (\ref{2.2}), the Szekeres models are covariant with
the transformations $r = f(r')$, where $f(r')$ is an arbitrary function.

The Szekeres metric has in general no symmetry, but acquires a 3-dimensional
symmetry group with 2-dimensional orbits when $A$, $B_1$, $B_2$ and $C$ are all
constant (that is, when $\nu,_r = 0$).

The sign of $g(r)$ determines the geometry of the 2-surfaces of constant $t$ and
$r$ (and the symmetry of the constant $A$, $B_1$, $B_2$ and $C$ limit). The
geometry of these surfaces is spherical, planar or hyperbolic (pseudo-spherical)
when $g > 0$, $g = 0$ or $g < 0$, respectively. With $A$, $B_1$, $B_2$ and $C$
being functions of $r$, the surfaces $r =$ const within a single space $t =$
const may have different geometries, i.e.\ they can be spheres in one part of
the space and the surfaces of constant negative curvature elsewhere, the
curvature being zero at the boundary.

The sign of $k(r)$ determines the type of evolution; with $k > 0 = \Lambda$ the
model expands away from an initial singularity and then recollapses to a final
singularity, with $k < 0 = \Lambda$ the model is either ever-expanding or
ever-collapsing, depending on the initial conditions; $k = 0$ is the
intermediate case corresponding to the `flat' Friedmann model ($k = 0$ can also
occur on a 3-surface as the boundary between a region with $k > 0$ and another
one with $k < 0$). The sign of $k(r)$ influences the sign of $g(r)$. Since
$1/h^2$ in (\ref{2.4}) must be non-negative,\footnote{$1/h(r)$ can be zero at
isolated points -- it is then either a coordinate singularity or a neck or belly
-- but not on open intervals.} we have the following: With $g > 0$ (spherical
geometry), all three types of evolution are allowed, with $g = 0$ (plane
geometry), $k$ must be non-positive (only parabolic or hyperbolic evolutions are
allowed), and with $g < 0$ (hyperbolic geometry), $k$ must be strictly negative,
so only the hyperbolic evolution is allowed.

The Friedmann limit follows when $\Phi (t,r) = \Phi_1(r) S(t)$. No further
specialization of the Szekeres functions is needed; the limiting Friedmann model
is represented in the little-known Goode -- Wainwright \cite{GoWa1982}
coordinates, see also Ref. \cite{Kras1997}.

The Szekeres models are subdivided according to the sign of $g(r)$ into the
quasi-spherical (with $g > 0$), quasi-plane ($g = 0$) and quasi-hyperbolic ones
($g < 0$). Despite suggestions to the contrary made in the literature, the
geometry of the latter two classes has not been investigated at all and is not
really understood; see Refs. \cite{HeKr2008} and \cite{Kras2008} for recent work
on their interpretation. Only the quasi-spherical model has been rather well
investigated, and found useful application in astrophysical cosmology. However,
including $g \leq 0$ in the redshift equations causes no complication, so we
consider here an arbitrary $g$.

The quasi-spherical model may be imagined as a generalisation of the L--T model
in which the spheres of constant mass are made non-concentric. The functions
$A(r)$, $B_1(r)$ and $B_2(r)$ determine how the centre of a sphere changes its
position in a space $t =$ const when the radius of the sphere is increased or
decreased \cite{HeKr2002}. Still, this is a rather simple geometry because all
the arbitrary functions depend on one variable, $r$.

It is often convenient to reparametrise the Szekeres metric as follows
\cite{Hell1996}. Even if $A = 0$ initially, a transformation of the $(x,
y)$-coordinates can restore $A \neq 0$, so we may assume $A \neq 0$ with no loss
of generality \cite{PlKr2006}. Then let $g \neq 0$. Writing $A =
\sqrt{|g|}/(2S)$, $B_1 = - \sqrt{|g|} P/(2S)$, $B_2 = - \sqrt{|g|} Q/(2S)$,
$\varepsilon \df g / |g|$, $k = |g| \widetilde{k}$ and $\Phi = \sqrt{|g|}
\widetilde{\Phi}$, we can represent the metric (\ref{2.2}) as\footnote{The
tildes were dropped in (\ref{2.7}) and in all further text. The $\Phi$ in
(\ref{2.7}) is in fact $\widetilde{\Phi}$ and the $k(r)$ is $\widetilde{k}(r)$.
The redefinitions imply, via (\ref{2.4}), $C = \sqrt{|g|}\left[\left(P^2 +
Q^2\right)/S + \varepsilon S\right]/ 2$, $h^2 = 1/[|g| (\varepsilon -
\widetilde{k})]$ and $M = \sqrt{|g|}^3 \widetilde{M}$. The $M$ used from now on
is in fact $\widetilde{M}$.
  }
\begin{eqnarray}\label{2.7}
{\rm e}^{- \nu} &=& \sqrt{|g|} {\cal{E}}, \nonumber \\
{\cal E} &\df& \frac  {(x - P)^2} {2S} + \frac {(y - Q)^2} {2S} + \frac
{\varepsilon S} 2, \\
{\rm d} s^2 &=& {\rm d} t^2 - \frac {\left(\Phi,_r - \Phi {\cal E},_r/{\cal
E}\right)^2} {\varepsilon - k(r)} {\rm d} r^2 - \frac {\Phi^2} {{\cal E}^2}
\left({\rm d} x^2 + {\rm d} y^2\right),\nonumber
\end{eqnarray}
where, so far, $\varepsilon = \pm 1$ ($+1$ for the quasi-spherical and $-1$ for
the quasi-hyperbolic model). When $g = 0$, the transition from (\ref{2.2}) to
(\ref{2.7}) is $A = 1/(2S)$, $B_1 = - P/(2S)$, $B_2 = - Q/(2S)$ and $\Phi$ is
unchanged.\footnote{The implied changes in $C$ and $h$ are then $C = (P^2 +
Q^2)/(2S)$, $h^2 = -1/k$; $k$ and $M$ remain unchanged.} Then (\ref{2.7})
applies with $\varepsilon = 0$, and the resulting model is quasi-plane.

The parametrisation introduced above makes several formulae simpler, mainly
because the constraint (\ref{2.4}) is identically fulfilled in it. However, this
parametrisation obscures the fact, evident in (\ref{2.1}) -- (\ref{2.4}), that
{\it the same} Szekeres model may be quasi-spherical in one part of the
spacetime, and quasi-hyperbolic elsewhere, with the boundary between these two
regions being quasi-plane; see an explicit simple example in Ref.
\cite{HeKr2008}. In most of the literature published so far, these models have
been considered separately, but this was either for purposes of systematic
research, or with a specific application in view that fixed the sign of $g(r)$.

Equation (\ref{2.3}), is formally identical to the Friedmann equation, but with
$k$ and $M$ depending on $r$, so each surface $r$ = const evolves independently
of the others. The solutions $\Phi(t, r)$ are the same as the corresponding L--T
solutions, and are unaffected by the dependence of the Szekeres metric on the
$(x,y)$ coordinates.

As defined by (\ref{2.2}) -- (\ref{2.3}), the Szekeres models contain 8
functions of $r$, of which only 7 are arbitrary because of (\ref{2.4}). The
parametrisation of (\ref{2.7}) turns
$g(r)$ to a constant parameter $\varepsilon$, thus reducing the number to 6. By
a choice of $r$ (still arbitrary up to now), we can fix one more function (for
example, by defining $r' = M(r)$). Thus, the number of arbitrary functions that
correspond to physical degrees of freedom is 5.

In the following, we will represent the Szekeres solutions with $\beta,_r \neq
0$ in the parametisation introduced in (\ref{2.7}). The formula for mass density
in these variables is
\begin{equation}\label{2.8}
\kappa \rho = \frac {2 \left(M,_r - 3 M {\cal E},_r / {\cal E}\right)} {\Phi^2
\left(\Phi,_r - \Phi {\cal E},_r / {\cal E}\right)}.
\end{equation}
The shear tensor is
\begin{equation}\label{2.9}
\sigma^{\alpha}{}_{\beta} = \frac{1}{3} \left( \frac{\Phi,_{tr} - \Phi,_t
\Phi,_r/ \Phi} {\Phi,_r - \Phi {\cal E},_r/{\cal E}} \right) {\rm diag}
(0,2,-1,-1),
\end{equation}
and the scalar of expansion is
\begin{equation}\label{2.10}
\theta = u^{\alpha}{}_{;\alpha} =  \frac {2 \Phi,_t} {\Phi} + \frac{\Phi,_{tr} -
\Phi,_t {\cal E},_r/{\cal E}} {\Phi,_r - \Phi {\cal E},_r/{\cal E}}.
\end{equation}

\section{Remarks on the Bondi redshift equation in the L--T model}\label{Bondi}

\setcounter{equation}{0}

The L--T model is a special case of the quasi-spherical Szekeres models that
follows from (\ref{2.7}) when $\varepsilon = +1$ and the functions $P$, $Q$, $S$
are all constant. With a different representation of the coordinates on a
sphere, the resulting metric is:
\begin{equation}\label{3.1}
{\rm d} s^2 = {\rm d} t^2 - \frac {{R,_r}^2} {1 + 2E(r)} {\rm d} r^2 - R^2(t, r)
\left({\rm d} \vartheta^2 + \sin^2 \vartheta {\rm d} \varphi^2\right),
\end{equation}
and the equation of an incoming radial null geodesic is
\begin{equation}\label{3.2}
\dr t r = - \frac {R,_r(t, r)} {\sqrt{1 + 2E}}.
\end{equation}
Bondi's derivation \cite{Bond1947} of the redshift equation for this geodesic is
as follows. Take a light signal obeying (\ref{3.2}), the equation of its
trajectory (the solution of (\ref{3.2})) is
\begin{equation}\label{3.3}
t = T(r)
\end{equation}
Take a second light signal, emitted from the same radial coordinate $r$, but
later (as measured by the time coordinate $t$) by $\tau$. The equation of its
trajectory is:
\begin{equation}\label{3.4}
t = T(r) + \tau(r),
\end{equation}
where $(T + \tau)$ obeys, from (\ref{3.2}):
\begin{equation}\label{3.5}
\dr T r + \dr {\tau} r = - \frac {R,_r(T(r) + \tau(r), r)} {\sqrt{1 + 2E(r)}}.
\end{equation}
{}From the Taylor formula we have:
\begin{eqnarray}\label{3.6}
R,_r(T(r) + \tau(r), r) &=& R,_r(T(r), r) + \tau(r) R,_{tr}(T(r), r) \nonumber
\\
&+& {\cal O}(\tau^2, r),
\end{eqnarray}
where the last term has the property ${\cal O}(\tau^2, r)/\tau \llim{\tau \to 0}
0$. Now, {\em assuming that $\tau$ is small}, we neglect the last term in
(\ref{3.6}) and obtain from (\ref{3.5}), taking into account (\ref{3.2}):
\begin{equation}\label{3.7}
\dr {\tau} r = - \tau(r) \frac {R,_{tr}(T(r), r)} {\sqrt{1 + 2E(r)}}.
\end{equation}
If $\tau$ is the period of an electromagnetic wave, then by definition:
\begin{equation}\label{3.8}
\frac {\tau(r_{\rm obs})} {\tau(r_{\rm em})} = 1 + z(r_{\rm em}),
\end{equation}
where the subscripts `obs' and `em' refer to the points of observation and
emission, respectively, and $z$ is the redshift. {}From (\ref{3.8}), keeping the
observer at a fixed position and letting $r_{\rm em}$ vary, we obtain $(\dril
{\tau} r) / \tau = - (\dril z r) / (1 + z)$, and so in (\ref{3.7}):
\begin{equation}\label{3.9}
\frac 1 {1 + z}\ \dr z r = \frac {R,_{tr}(T(r), r)} {\sqrt{1 + 2E(r)}}.
\end{equation}
This is Bondi's radial redshift equation \cite{Bond1947}. It does not describe
the redshift propagation exactly. Neglecting the last term in (\ref{3.6}) we
have changed the exact equation into one that only approximates the actual
variation of $\tau$ along the ray. The approximation is better the smaller the
value of $\tau$. Considering that $\tau$ is the period of an electromagnetic
wave, and taking into account the period range of relevance in observational
astronomy (from gamma rays up to radio waves, the longest observed of which have
the wavelength of the order of 15 m, thus the period of about $5 \times 10^{-8}$
s), we see that, compared to cosmological time-scales, the periods are short
indeed and the approximation is not bad. Moreover, as seen from (\ref{3.8}), by
following the rays back from the observation event into the past, we encounter
ever smaller values of $\tau$, so the approximation gets progressively better
with increasing redshift (or, rather, gets progressively worse as the ray
approaches us). Still, it is conceptually important to remember that (\ref{3.9})
involves an approximation (this approximation is equivalent to the geometric
optics approximation [11, 24] that leads to the commonly used expression for the
redshift $1 + z = (k_\alpha u^\alpha)_{\rm em} / (k_\alpha u^\alpha)_{\rm
obs}$).

We shall apply the same approach to the redshift equations in the
Szekeres models in Sec. \ref{Szredshift}.

\section{Equations of general null geodesics in a Szekeres
spacetime}\label{nullgeo}

\setcounter{equation}{0}

For reference, the equations of general null geodesics in a Szekeres model are
copied from Ref. \cite{BKHC2009} in Appendix \ref{nullaffine}.\footnote{It is
shown in Ref. \cite{BKHC2009} (see also Ref. \cite{NoDe2007}) that in general
there exists no analogue of a radial null geodesic. Radial geodesics exist only
when the Szekeres model is axially symmetric; then their intersections with
every space of constant time coordinate lie on the axis of symmetry.} They are
written there in terms of an affine parameter $s$. For our present purpose it is
more convenient to use the coordinate $r$ as an independent parameter (which is
non-affine).

This is allowed, but with some caution. It is easily seen from (\ref{a.1}) --
(\ref{a.4}) in Appendix \ref{nullaffine} that a geodesic on which $\dril r s =
0$ over some open range of $s$ has $\dril x s = \dril y s = 0$ in that range,
and so is timelike. However, (\ref{a.1}) -- (\ref{a.4}) do not guarantee that
$\dril r s \neq 0$ at all points; isolated points at which $\dril r s =0$ can
exist. Examples that explain how this can happen are the non-radial geodesics in
an L--T model, considered in Sec. \ref{numex}. Thus, $r$ can be used as a
parameter on null geodesics only on such segments where $\dril s r > 0$ or
$\dril s r < 0$ throughout.

Several sub-expressions in the equations of Appendix \ref{nullaffine} are
multiply repeated, therefore we introduce the following abbreviations:
\begin{eqnarray}
\Phi,_r - \Phi {\cal E},_r/{\cal E} &\df& \Phi_1, \label{4.1} \\
\Phi,_{tr} - {\Phi,_t} {\cal E},_r/{\cal E} &\df& \Phi_{01}, \label{4.2} \\
\Phi,_{rr} - \Phi {\cal E},_{rr}/{\cal E} &\df& \Phi_{11}, \label{4.3} \\
{\cal E},_r  {\cal E},_x - {\cal E} {\cal E},_{xr} &\df& E_{12}, \label{4.4} \\
{\cal E},_r  {\cal E},_y - {\cal E} {\cal E},_{yr} &\df& E_{13}. \label{4.5}
\end{eqnarray}
In addition, the following replacement will appear useful:
\begin{equation}
\left(\dr x r\right)^2 + \left(\dr y r\right)^2 \df \Sigma. \label{4.6}
\end{equation}

We have, for any coordinate:
\begin{equation}\label{4.7}
\dr {^2 x^\alpha} {s^2} = \left(\dr r s\right)^2 \dr {^2 x^\alpha} {r^2} + \dr
{^2 r} {s^2} \dr {x^\alpha} r.
\end{equation}
Then, from (\ref{a.2}) we have:
\begin{eqnarray}\label{4.8}
&& \frac {{\rm d}^2 r} {{\rm d} s^2} = \left(\dr r s\right)^2 \left\{- 2 \frac
{\Phi_{01}} {\Phi_1} \dr t r - \left(\frac{\Phi_{11}} {\Phi_1} - \frac {{\cal
E},_r} {\cal E} + \frac 1 2 \frac {k,_r} {\varepsilon - k}\right)\right.
\nonumber \\
&& - \left.2 \frac{\Phi}{{\cal E}^2} \frac{E_{12}} {\Phi_1} \dr x r - 2 \frac
{\Phi} {{\cal E}^2} \frac {E_{13}} {\Phi_1 } \dr y r + \frac {\Phi} {{\cal E}^2}
\frac {\varepsilon - k} {\Phi_1} \Sigma\right\} \nonumber \\
&& \df U(t, r, x, y) \left(\dr r s\right)^2.
\end{eqnarray}
Consequently, (\ref{a.1}), (\ref{a.3}) and (\ref{a.4}) become, using
(\ref{4.7}):
\begin{equation}\label{4.9}
\frac{{\rm d}^2 t}{{\rm d} r^2} + \frac {\Phi_1 \Phi_{01}} {\varepsilon - k} +
\frac{\Phi {\Phi,_t}}{{\cal E}^2} \Sigma + U \dr t r = 0,
\end{equation}
\begin{eqnarray}\label{4.10}
\frac {{\rm d}^2 x} {{\rm d} r^2} &+& 2 \frac {\Phi,_t} {\Phi} \dr t r \dr x r -
\frac 1 {\Phi} \frac {\Phi_1} {\varepsilon - k} E_{12} \nonumber \\
&+& \frac {2 \Phi_1} {\Phi}  \dr x r - \frac {{\cal E},_x} {\cal E} \left(\dr x
r\right)^2 - 2 \frac {{\cal E},_y} {{\cal E}} \dr x r \dr y r \nonumber \\
&+& \frac {{\cal E},_x} {\cal E} \left(\dr y r\right)^2 + U \dr x r = 0,
\end{eqnarray}
\begin{eqnarray}\label{4.11}
\frac {{\rm d}^2 y} {{\rm d} r^2} &+& 2 \frac {{\Phi,_t}} {\Phi} \dr t r \dr y r
- \frac 1 {\Phi} \frac {\Phi_1} {\varepsilon - k} E_{13} \nonumber \\
&+& \frac {2 \Phi_1} {\Phi}  \dr y r + \frac {{\cal E},_y} {\cal E} \left(\dr x
r\right)^2 - 2 \frac {{\cal E},_x} {\cal E} \dr x r \dr y r \nonumber \\
&-& \frac {{\cal E},_y} {\cal E} \left (\dr y r\right)^2 + U \dr y r = 0.
\end{eqnarray}

\section{The redshift equations in the Szekeres models}\label{Szredshift}

\setcounter{equation}{0}

Consider, in the Szekeres metric (\ref{2.7}), two light signals, the second one
following the first one after a short time-interval $\tau$, both emitted by the
same source and arriving at the same observer of coordinates $(r, x, y)$. The
equation of the trajectory of the first signal is
\begin{equation}\label{5.1}
(t, x, y) = (T(r), X(r), Y(r)),
\end{equation}
the corresponding equation for the second signal is
\begin{equation}\label{5.2}
(t, x, y) = (T(r) + \tau(r), X(r) + \zeta(r), Y(r) + \psi(r)).
\end{equation}
This means that while the first ray intersects the hypersurface of a given
constant value of the $r$-coordinate at the point $(t, x, y) = (T, X, Y)$, the
second ray intersects the same hypersurface at the point $(t, x, y) = (T + \tau,
X + \zeta, Y + \psi)$. Thus, in general, those two rays will not intersect the
same succession of intermediate matter worldlines on the way. Note that the
coordinates we use throughout the paper are comoving, so both the source of
light and the observer keep their spatial coordinates unchanged throughout
history. Given this, and given that we consider a pair of rays emitted by the
same source and received by the same observer, we have $(\zeta, \psi) = (0, 0)$
at the point of emission and at the point of reception. However, we have to
allow that the second ray was emitted in a different direction than the first
one, and is received from a different direction by the observer.\footnote{This
means that in a general inhomogeneous and anisotropic Universe the observed
objects should drift across the sky. See a brief quantitative discussion of this
effect in Sec. \ref{numex}.} The directions of the two rays will be determined
by $(\dril x r, \dril y r)$ and $(\dril x r + \xi(r), \dril y r + \eta(r))$,
respectively, where $\xi = \dril {\zeta} r$, $\eta = \dril {\psi} r$. We will
assume that $(\dril {\tau} r, \zeta, \psi, \xi, \eta)$ are small of the same
order as $\tau$, so we will neglect all terms nonlinear in any of them and terms
involving their products.

Since $\zeta = \psi = 0$ at the observer, these quantities are not in fact
observable. However, they have to be numerically monitored along the ray
because, as will be seen below, they enter the equation for $\tau$, which is
connected to the redshift by (\ref{3.8}).

In writing out the equations of propagation of redshift, we will introduce the
symbol $\Delta$. It will denote the difference between the relevant expression
taken at $(t + \tau, r, x + \zeta, y + \psi)$ and at $(t, r, x, y)$, linearized
in $(\tau, \zeta, \psi)$, for example $\Phi_{01}(t + \tau, r, x + \zeta, y +
\psi) - \Phi_{01}(t, r, x, y) \df \Delta \Phi_{01} + {\cal O}(\tau^2, \tau
\zeta, \tau \psi, \zeta^2, \dots)$. We have:\footnote{A quick way to calculate
(\ref{5.3}) -- (\ref{5.16}) is to take the differential of the corresponding
quantity at constant $r$ and replace $({\rm d} t, {\rm d} x, {\rm d} y, {\rm d}
(\dril x r), {\rm d} (\dril y r))$ by $(\tau, \zeta, \psi, \xi, \eta)$.}
\begin{eqnarray}
\Delta \Phi &=& \Phi,_t \tau, \qquad \Delta \left(\Phi,_t\right) = \Phi,_{tt}
\tau, \nonumber \\
\Delta \dr t r &=& \dr \tau r, \qquad \Delta \dr x r = \xi, \qquad \Delta \dr y
r = \eta, \label{5.3} \\
\Delta {\cal E} &=& {\cal E},_x \zeta + {\cal E},_y \psi, \nonumber \\
\Delta {\cal E},_x &=& \zeta/S, \qquad \Delta {\cal E},_y = \psi/S \label{5.4}
\\
\Delta \Phi_1 &=& \Phi_{01} \tau + \frac {\Phi E_{12}} {{\cal E}^2} \zeta +
\frac {\Phi E_{13}} {{\cal E}^2} \psi, \label{5.5} \\
\Delta \Phi_{01} &=& \left(\Phi,_{ttr} - \Phi_{tt} {\cal E},_r/{\cal E}\right)
\tau + \frac {\Phi,_t E_{12}} {{\cal E}^2} \zeta + \frac {\Phi,_t E_{13}} {{\cal
E}^2} \psi, \nonumber \\
\label{5.6} \\
\Delta \Phi_{11} &=& \left(\Phi,_{trr} - \Phi_{t} {\cal E},_{rr}/{\cal E}\right)
\tau + \frac {\Phi} {{\cal E}^2} \left({\cal E},_{rr} {\cal E},_x - {\cal E}
{\cal E},_{rrx}\right) \zeta \nonumber \\
&+& \frac {\Phi} {{\cal E}^2} \left({\cal E},_{rr} {\cal E},_y - {\cal E} {\cal
E},_{rry}\right) \psi. \label{5.7}
\end{eqnarray}
In the next two equations account is taken of the fact that ${\cal E},_{xy}
\equiv 0$.
\begin{eqnarray}
\Delta E_{12} &=& \left({\cal E},_r {\cal E},_{xx} - {\cal E} {\cal
E},_{rxx}\right) \zeta + \left({\cal E},_{ry} {\cal E},_x - {\cal E},_y {\cal
E},_{rx}\right) \psi, \nonumber \\
\label{5.8} \\
\Delta E_{13} &=& \left({\cal E},_{rx} {\cal E},_y - {\cal E},_x {\cal
E},_{ry}\right) \zeta + \left({\cal E},_r {\cal E},_{yy} - {\cal E} {\cal
E},_{ryy}\right) \psi, \nonumber \\
\label{5.9} \\
\Delta \Sigma &=& 2 \dr x r \xi + 2 \dr y r \eta, \label{5.10}
\end{eqnarray}
\begin{eqnarray}
\Delta U &=& 2 \left(- \frac {\Delta \Phi_{01}} {\Phi_1} + \frac {\Phi_{01}
\Delta \Phi_{1}} {{\Phi_1}^2}\right) \dr t r - 2 \frac {\Phi_{01}} {\Phi_1}\ \dr
{\tau} r \nonumber \\
&-& \frac {\Delta \Phi_{11}} {\Phi_1} + \frac {\Phi_{11} \Delta \Phi_{1}}
{{\Phi_1}^2} + \frac {\Delta {\cal E},_r} {\cal E} - \frac {{\cal E},_r \Delta
{\cal E}} {{\cal E}^2} \nonumber \\
&+& 2 \left(- \frac {\Phi,_t E_{12} \tau} {{\cal E}^2 \Phi_1} + 2 \frac {\Phi
\Delta {\cal E} E_{12}} {{\cal E}^3 \Phi_1} - \frac {\Phi \Delta E_{12}} {{\cal
E}^2 \Phi_1} \right. \nonumber \\
&\ & \ \ \ \ \ \ + \left.\frac {\Phi E_{12} \Delta \Phi_1} {{\cal E}^2
{\Phi_1}^2}\right) \dr x r - 2 \frac {\Phi E_{12} \xi} {{\cal E}^2 \Phi_1}
\nonumber \\
&+& 2 \left(- \frac {\Phi,_t E_{13} \tau} {{\cal E}^2 \Phi_1} + 2 \frac {\Phi
\Delta {\cal E} E_{13}} {{\cal E}^3 \Phi_1} - \frac {\Phi \Delta E_{13}} {{\cal
E}^2 \Phi_1} \right. \nonumber \\
&\ & \ \ \ \ \ \ + \left.\frac {\Phi E_{13} \Delta \Phi_1} {{\cal E}^2
{\Phi_1}^2}\right) \dr y r - 2 \frac {\Phi E_{13} \eta} {{\cal E}^2 \Phi_1}
\nonumber \\
&+& \frac {(\varepsilon - k) \Phi \Sigma} {{\cal E}^2 \Phi_1} \left(\frac
{\Phi,_t \tau} {\Phi} - 2 \frac {\Delta {\cal E}} {\cal E} - \frac {\Delta
\Phi_1} {\Phi_1} + \frac {\Delta \Sigma} {\Sigma}\right). \nonumber \\
 \label{5.11}
\end{eqnarray}

Applying the $\Delta$-operation to (\ref{4.9}) -- (\ref{4.11}) we obtain:
\begin{eqnarray}\label{5.12}
\dr {^2\tau} {r^2} &+& \frac {\Phi_{01} \Delta \Phi_{1} + \Phi_{1} \Delta
\Phi_{01}} {\varepsilon - k} + \frac {\left({\Phi,_t}^2 + \Phi \Phi,_{tt}\right)
\Sigma \tau} {{\cal E}^2} \nonumber \\
&-& 2 \frac {\Phi \Phi,_t \Delta {\cal E} \Sigma} {{\cal E}^3} + \frac {\Phi
\Phi,_t \Delta \Sigma} {{\cal E}^2} + \Delta U \dr t r + U \dr {\tau} r = 0,
\nonumber \\
\end{eqnarray}
\begin{eqnarray}\label{5.13}
\dr {^2\zeta} {r^2} &+& 2 \left(\frac {\Phi,_{tt}} {\Phi} - \frac {{\Phi,_t}^2}
{\Phi^2}\right) \dr t r \dr x r \tau + 2 \frac {\Phi,_t} {\Phi} \dr x r \dr
{\tau} r \nonumber \\
&+& 2 \frac {\Phi,_t} {\Phi} \dr t r \xi - \frac {\Delta \Phi_1 E_{12}}
{(\varepsilon - k) \Phi} + \frac {\Phi,_t \Phi_1 E_{12} \tau} {(\varepsilon - k)
\Phi^2} - \frac {\Phi_1 \Delta E_{12}} {(\varepsilon - k) \Phi} \nonumber \\
&+& 2 \left(\frac {\Delta \Phi_1} {\Phi} - \frac {\Phi_1 \Phi,_t \tau}
{\Phi^2}\right) \dr x r + 2 \frac {\Phi_1} {\Phi} \xi \nonumber \\
&-& \left(\dr x r\right)^2 \left(\frac {\zeta} {S {\cal E}} - \frac {{\cal E},_x
\Delta {\cal E}} {{\cal E}^2}\right) - \frac {2 {\cal E},_x
\xi} {\cal E} \dr x r \nonumber \\
&-& 2 \dr x r \dr y r \left(\frac {\psi} {S {\cal E}} - \frac {{\cal E},_y
\Delta {\cal E}} {{\cal E}^2}\right) - 2 \frac {{\cal E},_y} {\cal E} \left(\dr
y r \xi + \dr x r \eta\right) \nonumber \\
&+& \left(\dr y r\right)^2 \left(\frac {\zeta} {S {\cal E}} - \frac {{\cal E},_x
\Delta {\cal E}} {{\cal E}^2}\right) + \frac {2 {\cal E},_x \eta} {\cal E} \dr y
r \nonumber \\
&+& \Delta U \dr x r + U \xi = 0,
\end{eqnarray}
\begin{eqnarray}\label{5.14}
\dr {^2\psi} {r^2} &+& 2 \left(\frac {\Phi,_{tt}} {\Phi} - \frac {{\Phi,_t}^2}
{\Phi^2}\right) \dr t r \dr y r \tau + 2 \frac {\Phi,_t} {\Phi} \dr y r \dr
{\tau} r \nonumber \\
&+& 2 \frac {\Phi,_t} {\Phi} \dr t r \eta - \frac {\Delta \Phi_1 E_{13}}
{(\varepsilon - k) \Phi} + \frac {\Phi,_t \Phi_1 E_{13} \tau} {(\varepsilon - k)
\Phi^2} - \frac {\Phi_1 \Delta E_{13}} {(\varepsilon - k) \Phi} \nonumber \\
&+& 2 \left(\frac {\Delta \Phi_1} {\Phi} - \frac {\Phi_1 \Phi,_t \tau}
{\Phi^2}\right) \dr y r + 2 \frac {\Phi_1} {\Phi} \eta \nonumber \\
&+& \left(\dr x r\right)^2 \left(\frac {\psi} {S {\cal E}} - \frac {{\cal E},_y
\Delta {\cal E}} {{\cal E}^2}\right) + \frac {2 {\cal E},_y \xi} {\cal E} \dr x
r \nonumber \\
&-& 2 \dr x r \dr y r \left(\frac {\zeta} {S {\cal E}} - \frac {{\cal E},_x
\Delta {\cal E}} {{\cal E}^2}\right) - 2 \frac {{\cal E},_x} {\cal E} \left(\dr
y r \xi + \dr x r \eta\right) \nonumber \\
&-& \left(\dr y r\right)^2 \left(\frac {\psi} {S {\cal E}} - \frac {{\cal E},_y
\Delta {\cal E}} {{\cal E}^2}\right) - \frac {2 {\cal E},_y \eta} {\cal E} \dr y
r \nonumber \\
&+& \Delta U \dr y r + U \eta = 0,
\end{eqnarray}

In addition, we have the first integral of the geodesic equations (\ref{4.9}) --
(\ref{4.11}):
\begin{equation}\label{5.15}
\left(\dr t r\right)^2 =\frac {\left(\Phi_1\right)^2} {\varepsilon - k} + \frac
{\Phi^2} {{\cal E}^2} \left[\left(\dr x r\right)^2 + \left(\dr y
r\right)^2\right],
\end{equation}
Applying the $\Delta$-operation to this we get
\begin{eqnarray}\label{5.16}
\dr {\tau} r \dr t r &=& \frac {\Phi_1 \Delta \Phi_1} {\varepsilon - k}
\nonumber \\
&+& \left(\frac {\Phi \Phi,_t \tau} {{\cal E}^2} - \frac {\Phi^2 \Delta {\cal
E}} {{\cal E}^3}\right) \left[\left(\dr x r\right)^2 + \left(\dr y
r\right)^2\right]
\nonumber \\
&+& \frac {\Phi^2} {{\cal E}^2} \left(\dr x r \xi + \dr y r \eta\right).
\end{eqnarray}
Note: $\dril t r < 0$ for an incoming ray.

\section{Repeatable light paths}\label{repeat}

\setcounter{equation}{0}

As attested by (\ref{5.12}) -- (\ref{5.14}), in a generic Szekeres model two
light rays connecting a given source to a given observer at different instants
of emission do not proceed through the same succession of intermediate matter
particles. We will now investigate under what conditions this intermediate
succession is the same. This property will be called repeatable light paths
(RLP).

For a RLP we have
\begin{equation}\label{6.1}
\zeta = \psi = \xi = \eta = 0
\end{equation}
all along the ray. Then (\ref{5.12}) decouples  from (\ref{5.13}) --
(\ref{5.14}) and just determines $\tau$ (and, with it, the redshift), if the
null geodesic equations are solved first. Equations (\ref{5.13}) -- (\ref{5.14})
become then:
\begin{eqnarray}\label{6.2}
&& 2 \left(\frac {\Phi,_{tt}} {\Phi} - \frac {{\Phi,_t}^2} {\Phi^2}\right) \dr t
r \dr x r \tau + 2 \frac {\Phi,_t} {\Phi} \dr x r \dr {\tau} r \nonumber \\
&&- \frac {\Delta \Phi_1 E_{12}} {(\varepsilon - k) \Phi} + \frac {\Phi,_t
\Phi_1 E_{12} \tau} {(\varepsilon - k) \Phi^2} \nonumber \\
&&+ 2 \left(\frac {\Delta \Phi_1} {\Phi} - \frac {\Phi_1 \Phi,_t \tau}
{\Phi^2}\right) \dr x r + \Delta U \dr x r = 0,
\end{eqnarray}
\begin{eqnarray}\label{6.3}
&& 2 \left(\frac {\Phi,_{tt}} {\Phi} - \frac {{\Phi,_t}^2} {\Phi^2}\right) \dr t
r \dr y r \tau + 2 \frac {\Phi,_t} {\Phi} \dr y r \dr {\tau} r \nonumber \\
&&- \frac {\Delta \Phi_1 E_{13}} {(\varepsilon - k) \Phi} + \frac {\Phi,_t
\Phi_1 E_{13} \tau} {(\varepsilon - k) \Phi^2} \nonumber \\
&&+ 2 \left(\frac {\Delta \Phi_1} {\Phi} - \frac {\Phi_1 \Phi,_t \tau}
{\Phi^2}\right) \dr y r + \Delta U \dr y r = 0.
\end{eqnarray}

These equations can be understood in 2 ways:

1. As equations defining special Szekeres spacetimes in which {\em all} null
geodesics are RLPs.

2. As equations defining special null geodesics which are RLPs in
subcases of the Szekeres spacetimes.

In the first interpretation, (\ref{6.2}) -- (\ref{6.3}) should be identities in
the components of $\dril {x^{\alpha}} r$. They are polynomials of degree 3 in
these components, and since $(\dril t r)^2$ does not appear in them, the
constraint (\ref{5.15}) plays no role -- all powers of $\dril {x^{\alpha}} r$
that do appear are independent. Equating to zero the coefficient of $(\dril x
r)^3$ in (\ref{6.2}) (which arises inside $\Delta U$, within $\Sigma$), and
taking into account that $\Delta {\cal E} = \Delta \Sigma = 0$ when (\ref{6.1})
holds, we get
\begin{equation}\label{6.4}
\Psi \df \Phi,_{tr} - \Phi,_t \Phi,_r/\Phi = 0.
\end{equation}
The integral of this is $\Phi = S(t) f(r)$, where $S$ and $f$ are arbitrary
functions. It is seen from (\ref{2.9}) that this means zero shear, i.e. the
Friedmann limit. With (\ref{6.4}) fulfilled, (\ref{6.2}) and (\ref{6.3}) become
identities, and (\ref{4.9}) -- (\ref{4.11}) reduce to the equations of general
null geodesics in a Friedmann spacetime.\footnote{We recall, however, that the
Friedmann limit is represented in the Goode -- Wainwright \cite{GoWa1982}
coordinates (see the remark in para 4 after (\ref{2.6})). Consequently, all
equations representing the Friedmann model will look unfamiliar.} With the
observer placed at the origin, the geodesics become radial, $\dril x r = \dril y
r = 0$, and then (\ref{5.16}) becomes equivalent to the ordinary Robertson --
Walker redshift formula, $1 + z = S(t_o)/S(t_e)$. To verify this, some
calculations are needed, in which Ref. \cite{PlKr2006} may prove helpful.

Thus, we have proven the following:

{\em \bf Corollary 1:}

{\em The only spacetimes in the Szekeres family in which all null geodesics have
repeatable paths are the Friedmann models.}\footnote{This is one more piece of
evidence of how exceptional the Robertson -- Walker class of models is.}

In the second interpretation of (\ref{6.2}) -- (\ref{6.3}), we consider 2 cases:

{\bf A. The general case: $\dril x r \neq 0 \neq \dril y r$ everywhere.}

Then we multiply (\ref{6.2}) by $\dril y r$, (\ref{6.3}) by $\dril x r$ and
subtract the results. Disregarding the familiar case $\Psi = 0$ we get
\begin{equation}\label{6.5}
E_{12} \dr y r - E_{13} \dr x r = 0.
\end{equation}
This, together with (\ref{6.2}), (\ref{5.12}) and (\ref{4.9}) -- (\ref{4.11})
defines a certain subcase of the Szekeres model and a class of curves in it.
Since both the subcase and the class will turn out to be empty, but the
calculations proving it are rather elaborate, we present them in Appendix
\ref{emptygen}.

{\bf B. The special cases: $\dril x r = 0$ or $\dril y r = 0$.}

These two cases are equivalent under the coordinate transformation $(x, y) =
(y', x')$, so we consider only the first one. Again disregarding $\Psi = 0$, we
get from (\ref{6.2}) $E_{12} = 0$. Then, (\ref{4.10}) implies two possibilities:

{\bf Ba)} ${\cal E},_x = 0$.

This is possible only when $P$ is constant, and then the geodesic lies in the
subspace $x = P$. Equations (\ref{4.11}) and (\ref{6.3}) still have to be
obeyed, while (\ref{4.10}) and (\ref{6.2}) are fulfilled identically. The simple
coordinate transformation $x = x' + P$ has then the same effect as if $P = 0$
and $x = 0$ along the geodesic. We show in Appendix \ref{specB} that in this
case, apart from the axially symmetric subcase mentioned below, RLPs may exist
only when the Szekeres metric has a 3-dimensional symmetry group. Such
spacetimes are considered in Sec. \ref{LTredshift}.

{\bf Bb)} $\dril y r = 0$.

The case $\dril x r = \dril y r = 0$, $\varepsilon = +1$ was investigated in
detail in Ref. \cite{BKHC2009}. It turned out that this can happen only when the
Szekeres spacetime is axially symmetric, and then along only one sub-family of
null geodesics -- those that intersect each $t =$ constant space on the symmetry
axis. We show in Appendix \ref{axialRLP} that this result applies also with
$\varepsilon \leq 0$, and that other RLPs may exist only with higher symmetries.

\section{RLPs in the $G_3/S_2$ models}\label{LTredshift}

\setcounter{equation}{0}

The symbol $G_3/S_2$ denotes such models that have 3-dimensional symmetry groups
acting on 2-dimensional orbits \cite{PlKr2006}. They result from the general
$\beta' \neq 0$ Szekeres family when the functions $(P, Q, S)$ are all constant.
The symmetry of the model is then spherical when $\varepsilon = +1$ (this is the
L--T model), pseudospherical (also called hyperbolic) when $\varepsilon = -1$
and plane when $\varepsilon = 0$.

Using the $G_3$ symmetry, the origin of the $(x, y)$ coordinates at $(x, y) =
(P, Q)$ can be moved to any location on the $S_2$ surfaces. So let us consider
the $S_2$ on which the first light ray is emitted, and let us choose the origin
of $(x, y)$ at the position of the emitter. Thus, in (\ref{4.9}) -- (\ref{4.11})
the initial point of the earlier null geodesic will have the coordinates $(x, y)
= (P, Q)$, and, at this point, ${\cal E},_x = {\cal E},_y = 0$. In addition, the
isotropy subgroup of $G_3$, existing in each case at every point of the
manifold, allows us to rotate the $(x, y)$ coordinates, with no loss of
generality, so that the initial value of $\dril y r$ for our chosen geodesic is
zero, i.e. so that the ray is initially tangent to the $y =$ constant subspace.
Equation (\ref{4.11}) shows that with such initial conditions (and with ${\cal
E},_y = 0$ at the initial point) we have $\dril {^2y} {r^2} = 0$ initially, and
so $\dril {^2y} {r^2} = 0 = \dril y r$ all along the geodesic.

With coordinates chosen in such a way, equations (\ref{4.11}) and (\ref{6.3})
are fulfilled identically. However, (\ref{6.2}) is not an identity and reduces
to:
\begin{eqnarray}\label{7.1}
&& \dr x r \left[2 \left(\frac {\Phi,_{tt}} {\Phi} - \frac {{\Phi,_t}^2}
{\Phi^2}\right) \dr t r \tau + 2 \frac {\Phi,_t} {\Phi} \dr {\tau} r \right.
\nonumber \\
&&+ 2 \left(\frac {\Phi,_{tr}} {\Phi} - \frac {\Phi,_t \Phi,_r} {\Phi^2}\right)
\tau + 2 \left(- \frac {\Phi,_{ttr}} {\Phi,_r} + \frac {{\Phi,_{tr}}^2}
{{\Phi,_r}^2}\right) \dr t r \tau \nonumber \\
&&- 2 \frac {\Phi,_{tr}} {\Phi,_r} \dr {\tau} r - \frac {\Phi,_{trr}} {\Phi,_r}
\tau + \frac {\Phi,_{rr} \Phi,_{tr}} {{\Phi,_r}^2} \tau \nonumber \\
&& \left.+ \frac {(\varepsilon - k) \tau} {{\cal E}^2} \left(\dr x r\right)^2
\left(\frac {\Phi,_t} {\Phi,_r} - \frac {\Phi \Phi,_{tr}}
{{\Phi,_r}^2}\right)\right] \nonumber \\
&& \df \dr x r\ \chi = 0.
\end{eqnarray}
One solution of this is $\dril x r = 0$, which together with $\dril y r = 0$
defines a radial null geodesic. Then, (\ref{4.10}) -- (\ref{4.11}) are fulfilled
identically, while (\ref{5.15}) -- (\ref{5.16}), together with (\ref{4.1}) --
(\ref{4.2}) and (\ref{5.5}) reproduce the Bondi equation (\ref{3.7}) when
$\varepsilon = +1$. So, we found that in the $G_3/S_2$ models all radial null
geodesics are RLPs.\footnote{The null geodesics with $\dril x r = \dril y r = 0$
can properly be called radial only in the L--T model, where $\varepsilon = +1$.
What this condition means in the other two cases is not clear, so the term
``radial'' is used here only as a brief label.}

There would exist other RLPs in these models if $\chi$ in (\ref{7.1}) were zero
along any null geodesic -- possibly in some subcases of the models. It is shown
in Appendix \ref{RLPinG3S2} that this does not happen, so the radial null
geodesics are {\em the only} RLPs in these models.

\section{Numerical examples of non-RLPs in the L--T model}\label{numex}

\begin{figure}
\includegraphics[scale = 0.7]{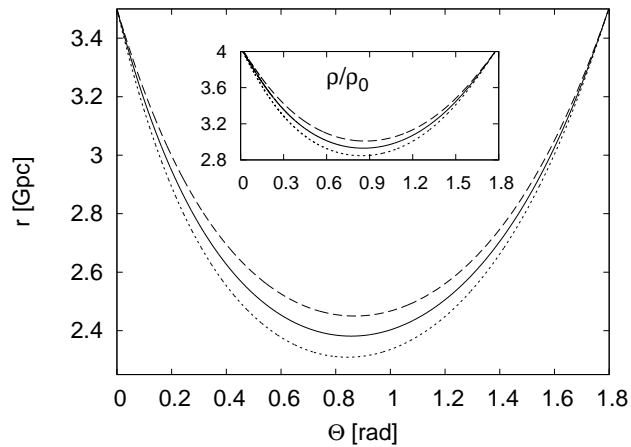}
\caption{Three nonradial null geodesics in an L--T model, projected on the space
$t =$ now along the flow lines of the L--T dust. Each geodesic runs between the
same observer and the same emitter, which, at present, lie at the same distance
of 3.5 Gpc from the center. The solid line represents the ray that the observer
receives at the current instant, $\gamma = 0.22 \pi$, the dashed line represents
the ray that was received $5 \times 10^9$ years ago, $\gamma = 0.228 \pi$, and
the dotted line represents the ray that will be received $5 \times 10^9$ years
in the future, $\gamma = 0.214 \pi$. As seen, nonradial null geodesics in the
L--T model do not have the RLP property. The inset shows the density profile
($\rho_0$ is the density at the origin) evaluated at the current instant along
these three different paths.} \label{fig1}
\end{figure}

For illustration, we first consider a configuration that is not realistic, but
shows the non-RLP effect in a clearly visible way. It is an LT model specified
by the following functions: $t_B =0$ and $\rho(t_0,r) =  \rho_0 \left[ 1 +
\delta - \delta \exp \left( - {r^2}/{\sigma^2} \right) \right]$, (where $t_0$ is
the current instant, $r$ is defined as $R(t_0,r)$, and $\rho_0$ is the density
at the origin and equals $0.3 \times (3H_0^2)/(8 \pi G)$, where $H_0 =
72$~km~s$^{-1}$~Mpc$^{-1}$ and $G$ is the gravitational constant). This model is
the so-called giant void model discussed in detail in \cite{BoWy2008}, with the
best-fit parameters: $\delta = 4.05$ and $\sigma = 2.96$ Gpc. We use this model
to study the configuration presented in Fig. \ref{fig1}, where, for the middle
curve, the angle between the radial direction and the incoming geodesic is
$\gamma = 0.22 \pi$. We consider 3 light paths. The first one corresponds to
photons received by the observer $5 \times 10^9$ years ago, the second one
corresponds to photons received at the current instant, and the third one
corresponds to photons which will be received in $5 \times 10^9$ year in the
future. Figure \ref{fig1} shows these 3 geodesics projected on the space $t =$
now along the flow lines of the matter source in the L--T model. Since in each
case the light paths are different, the profile of matter density along each
projected light ray is different. This feature is presented in the inset in Fig.
\ref{fig1}. Even though the density variation along the light path is of small
amplitude, the effect is clearly visible. The average rate of change of the
position of the source in the sky, seen by the observer, is $\sim 10^{-7}$ arc
sec per year.

\begin{figure}
\includegraphics[scale = 0.7]{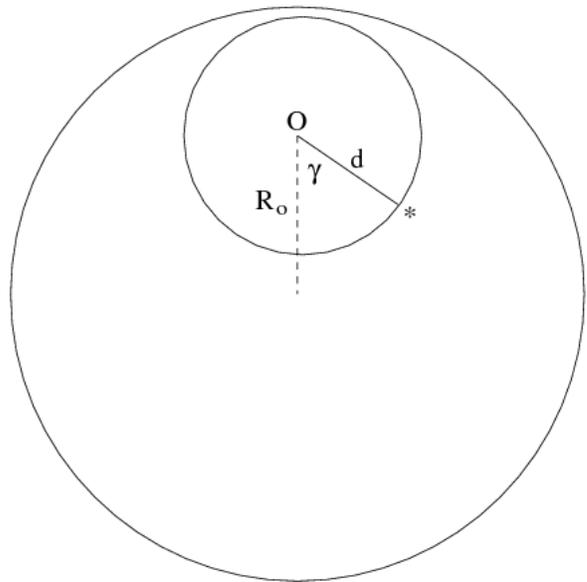}
\caption{A schematic view of the considered configurations. The observer (O) is
located at $R_0$ and observes a galaxy (*), the angle between the direction
towards the galaxy and towards the origin is $\gamma$. Because of the non-RLP
effect the angle at which the galaxy is observed at some other time instant is
different.} \label{fig2}
\end{figure}

Now we will study a more realistic configuration. The parameters of the L--T
model will be the same as above, but the placement of the observer and of the
source will be different, see Fig. \ref{fig2}. The observer (O) is located at
$R_0$ ($R_0 = R(t_0,r)$ is the present-day areal distance) and observes a galaxy
(*), the angle between the direction towards the galaxy and towards the origin
is $\gamma$. We study 3 configurations: (1) $R_0 = 3$ Gpc, (2) $R_0 = 1$ Gpc,
(3) $R_0 = 1$ Gpc but with $\delta = 10$. All 3 cases have $d = 1$ Gyr ($\approx
306.6$ Mpc). For each case (for a given $\gamma$) we find a null geodesic that
joins the observer and the galaxy. We then calculate the rate of change
$\gamma$, which is equivalent to the change of the position  of the galaxy in
the sky. A detailed description of the algorithm is presented in Appendix
\ref{ModAlg}. The results are presented in Fig. \ref{fig3}.

As seen, the rate of change of the position of the source in the sky depends on
the angle $\gamma$. The amplitude of the change is of the order $\sim 10^{-7}$
arc sec per year for case (2) and $\sim 10^{-6}$ arc sec per year for cases (1)
and (3). Given
Gaia\footnote{http://sci.esa.int/science-e/www/area/index.cfm?fareaid=26}
accuracy of position measurement, $5-20 \times 10^{-6}$ arc sec, we would need
to wait at least a few years to detect the change of position due to non-RLP
effects. However, this estimate assumes that we have a reference direction that
does not change. This will be a difficult practical problem, since cosmological
observations are done under the assumption that our Universe is precisely
represented in large scales by the Robertson -- Walker class of models, in which
there is no such drift. We would have to identify a direction that does not
change with time even in an inhomogeneous model or measure a relative change of
position between various objects.

\begin{figure}
\includegraphics[scale = 0.7]{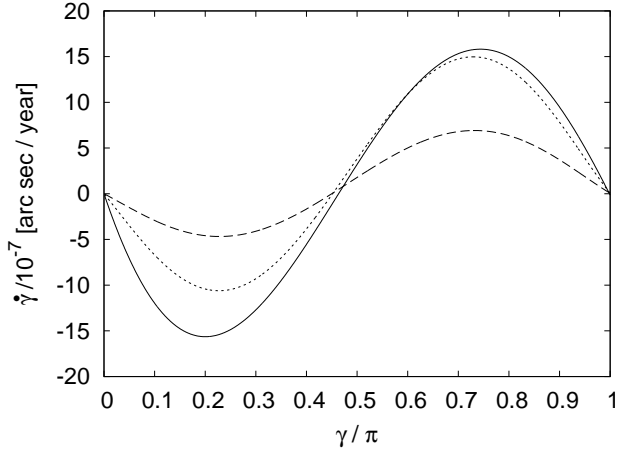}
\caption{ The rate of change of position in the sky ($\dot{\gamma}$) due to the
non-RLP effect, expressed as a change of an angle in arc sec per year $\times
10^{7}$. The solid  line presents case (1) where $R_0 = 3$ Gpc, the dashed line
presents case (2) where $R_0 = 1$ Gpc, and the dotted line presents case (3)
where $R_0 = 1$ Gpc and $\delta = 10$.}
 \label{fig3}
\end{figure}

\section{Summary}\label{summary}

By a method analogous to that of Bondi \cite{Bond1947} we have derived the
equations to be obeyed by the redshift in a general Szekeres $\beta' \neq 0$
spacetime, (\ref{5.12}) -- (\ref{5.14}). The null geodesic equations
parametrised by $r$, which must be solved together with (\ref{5.12}) --
(\ref{5.14}), are given by (\ref{4.9}) -- (\ref{4.11}). Although the physically
most interesting quantity is the longitudinal redshift determined by $\tau$, the
other two components, $\zeta$ and $\psi$, must be numerically monitored along
the ray because the equations that determine $(\tau, \zeta, \psi)$ are coupled.

We have shown that, in general, two light rays sent from the same source at
different times to the same observer do not proceed through the same succession
of intermediate matter particles; we refer to this property by saying that the
light paths are not repeatable. In a toy model, with the present spatial
distance between the light source B and the observer being of the order of 1.5
Gpc, the estimated rate of the drift of B across the sky would be $\approx 7
\times 10^{-8}$ arc sec per year. In a more realistic configuration, this number
is $\approx 10^{-6}$ arc sec per year. The Gaia is expected to have the
precision of position determination $5-20 \times 10^{-6}$ arc sec.

We have derived the equations defining repeatable light paths (RLPs),
(\ref{6.2}) -- (\ref{6.3}); they must hold together with (\ref{4.9}) --
(\ref{4.11}) and (\ref{5.12}). We have shown that all null geodesics are RLPs
only in the Friedmann models. The only other cases in which RLPs exist are the
following:

({\it i}) The axially symmetric Szekeres models, in which the RLPs are the null
geodesics intersecting every space of constant time on the axis of symmetry.

({\it ii}) The radial null geodesics in the $G_3/S_2$ subcases (i.e. in the
spacetimes that have 3-dimensional symmetry groups).

\appendix

\section{Equations of null geodesics in a Szekeres spacetime in an affine
parametrisation}\label{nullaffine}

For convenience of the readers, the equations of null geodesics in a Szekeres
spacetime are copied here from Ref. \cite{BKHC2009}. They are given in an affine
parametrisation.

\begin{eqnarray}
&& \frac {{\rm d}^2 t} {{\rm d} s^2} + \frac {\Phi,_{tr} - {\Phi,_t} {\cal
E},_r/{\cal E}} {\varepsilon - k} (\Phi,_r - \Phi {\cal E},_r/{\cal E})
\left(\dr r s\right)^2 \nonumber \\
&& + \frac {\Phi \Phi,_t} {{\cal E}^2} \left[\left(\dr x s\right)^2  + \left(\dr
y s\right)^2\right] = 0, \label{a.1} \\
&& \frac {{\rm d}^2 r} {{\rm d} s^2} + 2 \frac {{\Phi,_{tr}} - {\Phi,_t}{\cal
E},_r / {\cal E}} {\Phi,_r - \Phi {\cal E},_r/{\cal E}} \dr t s \dr r s \nonumber \\
&+& \left(\frac {\Phi,_{rr} - \Phi {\cal E},_{rr}/{\cal E}} {\Phi,_r - \Phi
{\cal E},_r/{\cal E}} - \frac {{\cal E},_r} {\cal E} + \frac 1 2 \frac {k,_r}
{\varepsilon - k}\right) \left(\dr r s\right)^2 \nonumber \\
&+& 2 \frac {\Phi} {{\cal E}^2} \frac {{\cal E},_r  {\cal E},_x - {\cal E} {\cal
E},_{xr}} {\Phi,_r - \Phi {\cal E},_r/{\cal E}} \dr r s \dr x s \nonumber \\
&+& 2 \frac {\Phi} {{\cal E}^2} \frac {({\cal E},_r {\cal E},_y - {\cal E} {\cal
E},_{yr})} {\Phi,_r - \Phi {\cal E},_r/{\cal E}} \dr r s \dr y s \nonumber \\
&-& \frac {\Phi} {{\cal E}^2} \frac {\varepsilon - k} {\Phi,_r - \Phi {\cal
E},_r/{\cal E}} \left[\left(\dr x s\right)^2 + \left(\dr y s\right)^2\right] =
0, \nonumber \\ \label{a.2} \\
&& \frac {{\rm d}^2 x} {{\rm d} s^2} + 2 \frac {\Phi,_t} {\Phi} \dr t s \dr x s
\nonumber \\
&-& \frac 1 {\Phi} \frac {{\Phi},_r - {\Phi} {\cal E},_r/{\cal E}} {\varepsilon
- k} ({\cal E},_r {\cal E},_x - {\cal E}  {\cal E},_{xr})
\left(\dr r s\right)^2 \nonumber \\
&+& \frac 2 {\Phi} \left(\Phi,_r - \Phi \frac {{\cal E},_r} {\cal E}\right) \dr
r s \dr x s - \frac {{\cal E},_x} {\cal E} \left(\dr x s\right)^2 \nonumber \\
&-& 2 \frac {{\cal E},_y} {\cal E} \dr x s \dr y s + \frac {{\cal E},_x} {\cal
E} \left(\dr y s\right)^2 = 0, \label{a.3} \\
&& \frac {{\rm d}^2 y} {{\rm d} s^2} + 2 \frac {\Phi,_t} {\Phi} \dr t s \dr y
s \nonumber \\
&-& \frac 1 {\Phi} \frac {\Phi,_r - \Phi {\cal E},_r/{\cal E}} {\varepsilon - k}
({\cal E},_r {\cal E},_y - {\cal E} {\cal E},_{yr}) \left(\dr r s\right)^2
\nonumber \\
&+& \frac 2 {\Phi} \left(\Phi,_r - \Phi \frac {{\cal E},_r} {\cal E}\right) \dr
r s \dr y s + \frac {{\cal E},_y} {\cal E} \left(\dr x s\right)^2 \nonumber \\
&-& 2 \frac {{\cal E},_x} {\cal E} \dr x s \dr y s - \frac {{\cal E},_y} {\cal
E} \left(\dr y s\right)^2 = 0. \label{a.4}
\end{eqnarray}

\section{Solutions of (\ref{6.5}).}\label{emptygen}

\setcounter{equation}{0}

Since (\ref{6.5}) should hold along certain null geodesics, its derivative by
$r$ along those geodesics must be zero. This derivative, denoted by ${\cal D} /
{\rm d} r$, of any quantity $\chi$ defined along the geodesic, $\chi(t(r), r,
x(r), y(r))$, is:
\begin{equation}\label{b.1}
\frac {{\cal D} \chi} {{\rm d} r} = \pdr {\chi} t\ \dr t r + \pdr {\chi} r +
\pdr {\chi} x\ \dr x r + \pdr {\chi} y\ \dr y r.
\end{equation}
Calculating ${\cal D} / {\rm d} r$ of (\ref{6.5}) we get:
\begin{eqnarray}\label{b.2}
&& \left(E_{12,r} + E_{12,x} \dr x r + E_{12,y} \dr y r\right) \dr y r \nonumber
\\
&& - \left(E_{13,r} + E_{13,x} \dr x r + E_{13,y} \dr y r\right) \dr x r
\nonumber \\
&& + E_{12} \dr {^2y} {r^2} - E_{13} \dr {^2 x} {r^2} = 0.
\end{eqnarray}
The expression in the last line can be calculated from (\ref{4.10}) --
(\ref{4.11}) using (\ref{6.5}); it is:
\begin{eqnarray}\label{b.3}
&& E_{12} \dr {^2y} {r^2} - E_{13} \dr {^2 x} {r^2} \nonumber \\
&& = \left({\cal E},_y {\cal E},_{rx} - {\cal E},_x {\cal E},_{ry}\right)
\left[\left(\dr x r\right)^2 + \left(\dr y r\right)^2\right].
\end{eqnarray}
Substituting (\ref{b.3}) and (\ref{4.4}) -- (\ref{4.5}) in (\ref{b.2}), and
taking into account the identities ${\cal E},_{xy} = {\cal E},_{xx} - {\cal
E},_{yy} = 0$ we get:
\begin{equation}\label{b.4}
E_{12,r} \dr y r - E_{13,r} \dr x r = 0.
\end{equation}
This should hold simultaneously with (\ref{6.5}). Since we assumed $\dril x r
\neq 0 \neq \dril y r$, (\ref{6.5}) and (\ref{b.4}) imply:
\begin{equation}\label{b.5}
E_{12} E_{13,r} - E_{13} E_{12,r} = 0.
\end{equation}
When (\ref{4.4}) -- (\ref{4.5}) are substituted in (\ref{b.5}), ${\cal E}$
factors out, and the other factor is:
\begin{eqnarray}\label{b.6}
&& {\cal E},_r \left({\cal E},_{y} {\cal E},_{rrx} - {\cal E},_{x} {\cal
E},_{rry}\right) \nonumber \\
&+& {\cal E},_{rr} \left({\cal E},_{x} {\cal E},_{ry} - {\cal E},_{y} {\cal
E},_{rx}\right) \nonumber \\
&+& {\cal E} \left({\cal E},_{rx} {\cal E},_{rry} - {\cal E},_{ry} {\cal
E},_{rrx}\right) = 0.
\end{eqnarray}
This simplifies to a polynomial of second degree in $x$ and $y$, which should
vanish identically.\footnote{Should there exist any point ${\cal P}_n$ in the
spacetime at which the polynomial would be nonzero, this would mean that the
determinant of the set $\{(\ref{6.5}), (\ref{b.4})\}$ is nonzero at ${\cal
P}_n$, which in turn would mean $\dril x r = \dril y r = 0$ at ${\cal P}_n$ --
contrary to our initial assumption.} Using the algebraic program Ortocartan
\cite{Kras2001, KrPe2000} we find that the coefficient of $(x^2 + y^2)$ is
\begin{equation}\label{b.7}
P,_{rr} Q,_r - P,_r Q,_{rr} = 0.
\end{equation}
One of the solutions of this is $P,_r = 0$; then no limitation for $Q$ follows.
This case we consider separately below.

When $P,_r \neq 0$, (\ref{b.7}) implies
\begin{equation}\label{b.8}
Q = C_0 P + D_0,
\end{equation}
where $C_0$ and $D_0$ are arbitrary constants. When this is substituted in
(\ref{b.6}), the coefficient of $y$ implies:
\begin{equation}\label{b.9}
\varepsilon \left(SS,_r P,_{rr} - {S,_r}^2 P,_r - SS,_{rr} P,_r\right) - \left(1
+ {C_0}^2\right) {P,_r}^3 = 0,
\end{equation}
and this guarantees that the whole of (\ref{b.6}) is fulfilled.

The case $\varepsilon = 0$ is seen to be incompatible with $P,_r \neq 0$. This
means that no RLPs exist in the $\varepsilon = 0$ models with $P,_r \neq 0$.
Further calculations apply only to $\varepsilon = \pm 1$.

In integrating (\ref{b.9}) we can assume $S,_r \neq 0$ because $S,_r = 0$
immediately implies $P,_r = 0$, which we have left for a separate investigation.
Therefore we can introduce $S(r)$ as the new independent variable in
(\ref{b.9}), which then becomes:
\begin{equation}\label{b.10}
\varepsilon \left(S P,_{SS} - P,_S\right) - \left(1 + {C_0}^2\right) {P,_S}^3 =
0.
\end{equation}
Since the case $P =$ constant was left for later, we assume $P,_S \neq 0$, and
then (\ref{b.10}) is easily integrated with the result:
\begin{equation}\label{b.11}
\varepsilon S^2 + \left(1 + {C_0}^2\right) P^2 = C_3 P + D_3,
\end{equation}
where $C_3$ and $D_3$ are new arbitrary constants.

When $\varepsilon = +1$, eqs. (\ref{b.8}) and (\ref{b.11}) are equivalent to
those that were shown in Ref. \cite{BKHC2009} (sec. 3.3.1) to be sufficient
conditions for the Szekeres metric to be axially symmetric. However, this
equivalence is nontrivial, and the extension of the proof to $\varepsilon = 0,
-1$ is not automatic, so we have to elaborate on this subject.

For this purpose, we note the following properties of the general Szekeres
metrics (\ref{2.7}):

(1) The metric (\ref{2.7}) does not change in form under the coordinate
transformation:
\begin{equation}\label{b.12}
(x, y) = (x' + x_0, y' + y_0),
\end{equation}
where $(x_0, y_0)$ are arbitrary constants. This changes $(P, Q)$ to
\begin{equation}\label{b.13}
(\widetilde{P}, \widetilde{Q}) = (P - x_0, Q - y_0).
\end{equation}

(2) The metric (\ref{2.7}) does not change in form when $(x, y)$ are transformed
by a general orthogonal transformation:
\begin{equation}\label{b.14}
x = \frac {ax' + by'} {\sqrt{a^2 + b^2}}, \qquad y = \frac {- bx' + ay'}
{\sqrt{a^2 + b^2}},
\end{equation}
which implies the change of $(P, Q)$ to:
\begin{equation}\label{b.15}
\widetilde{P} = \frac {aP - bQ} {\sqrt{a^2 + b^2}}, \qquad \widetilde{Q} = \frac
{bP + aQ} {\sqrt{a^2 + b^2}}.
\end{equation}

(3) The metric (\ref{2.7}) does not change in form under the discrete
transformations:
\begin{eqnarray}\label{b.16}
&& (x, y) = (y', x'), \quad (x, y) = (- x', y'), \nonumber \\
&& (x, y) = (x', - y'),
\end{eqnarray}
which induce, respectively
\begin{eqnarray}\label{b.17}
&& (\widetilde{P}, \widetilde{Q}) = (Q, P), \quad (\widetilde{P}, \widetilde{Q})
= (- P, Q), \nonumber \\
&& (\widetilde{P}, \widetilde{Q}) = (P, - Q).
\end{eqnarray}

(4) The metric (\ref{2.7}) does not change in form when $(x, y)$ are transformed
by a conformal symmetry of a Euclidean 2-plane -- a 2-dimensional Haantjes
transformation by the terminology of Ref. \cite{PlKr2006}. It has the form:
\begin{eqnarray}\label{b.18}
x &=& \frac {x' + \lambda_1 \left({x'}^2 + {y'}^2\right)} T, \nonumber \\
y &=& \frac {y' + \lambda_2 \left({x'}^2 + {y'}^2\right)} T, \\
T &\df& 1 + 2 \lambda_1 x' + 2 \lambda_2 y' + \left({\lambda_1}^2 +
{\lambda_2}^2\right) \left({x'}^2 + {y'}^2\right), \nonumber
\end{eqnarray}
where $\lambda_1$ and $\lambda_2$ are arbitrary constants -- the group
parameters. This group is Abelian, the inverse transformation to (\ref{b.18})
being of the same form, but with parameters $(- \lambda_1, -\lambda_2)$. The
characteristic properties of (\ref{b.18}), useful in calculations, are:
\begin{eqnarray}\label{b.19}
x^2 + y^2 &=& \frac {{x'}^2 + {y'}^2} T, \nonumber \\
{\rm d} x^2 + {\rm d} y^2 &=& \frac {{\rm d} {x'}^2 + {\rm d} {y'}^2} {T^2}.
\end{eqnarray}
Under (\ref{b.18}) -- (\ref{b.19}), $(P, Q, S)$ change, respectively, to:
\begin{eqnarray}\label{b.20}
&& \widetilde{P} = \frac 1 U \left[P - \lambda_1 \left(P^2 + Q^2 + \varepsilon
S^2\right)\right], \nonumber \\
&& \widetilde{Q} = \frac 1 U \left[Q - \lambda_2 \left(P^2 + Q^2 + \varepsilon
S^2\right)\right], \nonumber \\
&& \widetilde{S} = S / U, \nonumber \\
&& U \df 1 - 2 \lambda_1P - 2 \lambda_2Q \nonumber \\
&& \ \ \ \ \ \ \ \ \ + \left({\lambda_1}^2 + {\lambda_2}^2\right) \left(P^2 +
Q^2 + \varepsilon S^2\right).\ \ \ \ \ \ \
\end{eqnarray}

Let $\widetilde{\cal E}$ denote ${\cal E}$ with $(x, y, P, Q, S)$ replaced by
$(x', y', \widetilde{P}, \widetilde{Q}, \widetilde{S})$. Then calculation shows
that
\begin{equation}\label{b.21}
{\cal E} = {\widetilde{\cal E}} / T,
\end{equation}
and since $T$ does not depend on $r$, it follows that the $g_{rr}$ component in
(\ref{2.7}) is also covariant with (\ref{b.18}) -- (\ref{b.19}).

Now we will use the properties listed above to interpret the consequences of
(\ref{b.8}) and (\ref{b.11}) for the metric (\ref{2.7}).

The $D_0$ in (\ref{b.8}) can be set to zero by (\ref{b.12}) with $(x_0, y_0) =
(0, D_0)$. The $C_0$ in (\ref{b.8}) can be set to zero by (\ref{b.15}) with $b =
- a C_0$; the result of these two transformations is $Q = 0$. Finally, the $C_3$
in (\ref{b.11}) (with $C_0 = 0$ taken into account) can be set to zero by
(\ref{b.12}) with $(x_0, y_0) = (- C_3 / 2, 0)$. Thus we can assume $D_0 = C_0 =
C_3 = 0$ with no loss of generality.

We carry out a combination of (\ref{b.12}) with (\ref{b.18}):
\begin{eqnarray}\label{b.22}
x &=& x_0 + \frac {x' + \lambda_1 \left({x'}^2 + {y'}^2\right)} T, \nonumber \\
y &=& y',
\end{eqnarray}
and get the following generalisation of (\ref{b.20}) with $\lambda_2 = 0$:
\begin{eqnarray}\label{b.23}
&& \hspace{-4mm} \widetilde{P} = \frac 1 {\cal U} \left\{P - x_0 - \lambda_1
\left[\left(P - x_0\right)^2 + Q^2 + \varepsilon S^2\right]\right\}, \nonumber
\\
&& \hspace{-4mm} (\widetilde{Q}, \widetilde{S}) = (Q, S) / {\cal U}, \\
&& \hspace{-4mm} {\cal U} \df 1 - 2 \lambda_1 (P - x_0) + {\lambda_1}^2
\left[\left(P - x_0\right)^2 + Q^2 + \varepsilon S^2\right]. \nonumber
\end{eqnarray}
Using (\ref{b.8}) and (\ref{b.11}) with $D_0 = C_0 = C_3 = 0$, the above becomes
\begin{eqnarray}\label{b.24}
&& \hspace{-4mm} \widetilde{P} = \frac 1 {\cal U} \left[P - x_0 - \lambda_1
\left(-2 x_0 P + D_3 + {x_0}^2\right)\right],
\nonumber \\
&& \hspace{-4mm} \widetilde{Q} = 0, \\
&& \hspace{-4mm} {\cal U} \df 1 - 2 \lambda_1 (P - x_0) + {\lambda_1}^2 \left(-2
x_0 P + D_3 + {x_0}^2\right). \nonumber
\end{eqnarray}
Now it can be seen that if the constants $(x_0, \lambda_1)$, so far arbitrary,
obey:
\begin{eqnarray}\label{b.25}
&& 1 + 2 \lambda_1 x_0 = 0, \nonumber \\
&& x_0 + \lambda_1 \left(D_3 + {x_0}^2\right) = 0,
\end{eqnarray}
then $\widetilde{P} = \widetilde{Q} = 0$, and in the $(x', y')$ coordinates the
Szekeres metric is explicitly axially symmetric. However, two things must be
noted:

(1) The set (\ref{b.25}) has no solutions when $D_3 \leq 0$.

(2) With $P = Q = 0$, eq. (\ref{b.9}) is fulfilled identically, and (\ref{b.11})
no longer follows, thus there is no limitation on $S$.

Looking at (\ref{b.11}) with $C_0 = C_3 = 0$ we see that $D_3 < 0$ cannot occur
when $\varepsilon = +1$ or $\varepsilon = 0$. The case $D_3 = 0$, although
possible with $\varepsilon = +1$ or $\varepsilon = 0$, need not be considered
with these two values of $\varepsilon$, for the following reasons: With
$\varepsilon = +1$ this would imply $S = 0$, which is an impossibility in
(\ref{2.7}), and with $\varepsilon = 0 = C_0 = C_3$, (\ref{b.11}) implies $P =
0$. With $Q = 0$ now being considered, $P = \varepsilon = 0$ guarantees that $S$
may be set to 1 by a suitable reparametrisation of the other metric functions
\cite{HeKr2008}. Consequently, with $P = Q = 0$, the $\varepsilon = 0$ Szekeres
metric is already plane symmetric even with non-constant $S$, and the Szekeres
metrics with 3-dimensional symmetry groups are considered in Sec.
\ref{LTredshift}.

So, finally, $D_3 \leq 0$ must be considered only for $\varepsilon = -1$.
Since these calculations are lengthy and very complicated, we have moved them to
the separate appendix \ref{specmetric}.

We now come back to (\ref{b.7}) to consider the case $P,_r = 0$. By a
transformation of $x$ this can be reduced to $P = 0$. Then, the whole of
(\ref{b.6}) becomes:
\begin{equation}\label{b.26}
x \left[\varepsilon \left({S,_r}^2 Q,_r - SS,_r Q,_{rr} + SS,_{rr} Q,_r\right) +
{Q,_r}^3\right] = 0.
\end{equation}
This is equivalent to the subcase $C_0 = 0$ of (\ref{b.9}) under the coordinate
transformation $(x, y) = (\widetilde{y}, \widetilde{x})$ and the associated
renaming $(P, Q) = (\widetilde{Q}, \widetilde{P})$. This case was included in
the consideration above.

Thus, apart from the special cases $D_3 \leq 0$ to be considered further on,
RLPs with $\dril x r \neq 0 \neq \dril y r$ may possibly exist only when the
Szekeres metric is reducible, by a coordinate transformation, to one with $P = Q
= 0$. In this case, (\ref{6.5}) becomes:
\begin{equation}\label{b.27}
\varepsilon \frac {S,_r} S \left(x \dr y r - y \dr x r\right) = 0.
\end{equation}
But with $\varepsilon = 0$ and $P = Q = 0$ now being considered, the quasi-plane
Szekeres metric is plane symmetric even with non-constant $S$; see the paragraph
following (\ref{b.25}). The Szekeres metrics with 3-dimensional symmetry groups
are considered in Sec. \ref{LTredshift}, so we need not consider $\varepsilon =
0$ here.

When $S,_r = 0 = P,_r = Q,_r$, all Szekeres metrics acquire a 3-dimensional
symmetry group and are considered in Sec. \ref{LTredshift}. Thus, we need not
consider $S,_r = 0$ in (\ref{b.27}).

What remains of (\ref{b.27}) is $x \dril y r - y \dril x r = 0$. One solution of
this is $x = 0$ along the null geodesic. The other solution is $y = G_0 x$ along
the geodesic, where $G_0$ is a constant. However, we are now considering the
axially symmetric Szekeres solutions in which $P = Q = 0$. In this case, a
rotation (\ref{b.14}) can be chosen so that $y = G_0 x$ is transformed to $x' =
0$. So, apart from the special case $\varepsilon = -1, D_3 \leq 0$, for the
other Szekeres solutions the following result applies:

\medskip

{\bf Corollary 2:}

\medskip

The Szekeres spacetimes in which RLPs exist with $\dril x r \neq 0 \neq \dril y
r$ either are the Friedmann models (in which all null geodesics are RLPs) or are
inhomogeneous and axially symmetric or have a 3-dimensional symmetry group. In
the first two cases, coordinates may be chosen so that $x = 0$ along the
hypothetic RLP and $P = Q = 0$ in the metric.
The second case is considered in Appendix \ref{axialRLP}.
The third case is considered in Sec. \ref{LTredshift}
and in Appendix \ref{RLPinG3S2}.

\section{The special metric with $Q = 0$, $\varepsilon = -1$ and $D_3 \leq
0$.}\label{specmetric}

\setcounter{equation}{0}

We consider here the special case $D_3 \leq 0$ that arose in solving
(\ref{b.25}). The Szekeres model in question has
\begin{eqnarray}\label{c.1}
{\cal E} &=& \frac {x^2 - 2 P x + y^2 + D_3} {2S}, \nonumber \\
E_{12} &=& \frac {P,_r} {2 S^2} \left(y^2 - x^2 + D_3\right), \qquad E_{13} = -
\frac {P,_r} {S^2}\ xy.  \nonumber \\
\end{eqnarray}
The solution of (\ref{6.5}) is then either $P,_r = 0$, which belongs to the
axially symmetric case considered in appendix \ref{specB}, or
\begin{equation}\label{c.2}
x^2 + y^2 - Cy - D_3 = 0,
\end{equation}
where $C$ is the arbitrary constant that arises while integrating (\ref{6.5}).
By writing the above as $x^2 + (y - C/2)^2 = D_3 + C^2/4$ we note that the
following must hold:
\begin{equation}\label{c.3}
D_3 + C^2/4 > 0.
\end{equation}
(With this quantity being negative, (\ref{c.2}) has no solutions, i.e. there are
no RLPs. When it is zero, the only solution of (\ref{c.2}) is $(x, y) = (0,
C/2)$, what is possible only in the axially symmetric case of appendix
\ref{specB}.)

Note that (\ref{c.3}) implies $C \neq 0$, since $D_3 \leq 0$.

Taking the second derivative of (\ref{c.2}) by $r$ and substituting in it the
expressions for $x,_{rr}$ and $y,_{rr}$ from (\ref{4.10}) -- (\ref{4.11}), we
obtain an identity. This means that (\ref{c.2}) is consistent with the geodesic
equations (\ref{4.9}) -- (\ref{4.11}) and defines a special class of null
geodesics. We will verify in the following that this class does not contain any
RLPs.

We note the following auxiliary formulae. In Eqs. (\ref{c.4}) -- (\ref{c.11})
asterisks mark those equations that hold only along the null geodesics obeying
(\ref{c.2}), those without the asterisk are general. After (\ref{c.11}) all
further calculations are done only along these geodesics, so we omit the
asterisks for better readability.
\begin{eqnarray}
(*) && {\cal E} = \frac {- 2 P x + Cy + 2 D_3} {2S}, \label{c.4} \\
&& {\cal E},_x = \frac {x - P} S, \quad {\cal E},_y = \frac y S, \quad {\cal
E},_r = - \frac {x P,_r} S - \frac {S,_r} S\ {\cal E}, \nonumber \\
   \label{c.5} \\
&& {\cal E},_{rx} = - \frac {P,_r} S - \frac {S,_r} {S^2}\  (x - P), \quad {\cal
E},_{ry} = - \frac {y S,_r} {S^2}, \label{c.6} \\
(*) && E_{12} = \frac {P,_r} {2 S^2} \left(2 y^2 - Cy\right), \label{c.7}
\\
(*) && y,_r = - \frac {2x} {2y - C}\ x,_r, \label{c.8} \\
(*) && {x,_r}^2 + {y,_r}^2 = \frac {4D_3 + C^ 2} {(2y - C)^2}\ {x,_r}^2.
\label{c.9} \\
(*) && 2 {x,_r}^2 E_{12} + 2 x,_r y,_r E_{13} \nonumber \\
&& \ \ \ \ \ \ = \frac {P,_r} {S^2}\ y (2y - C) \left({x,_r}^2 +
{y,_r}^2\right). \label{c.10}
\end{eqnarray}
{}From (\ref{5.15}) we have
\begin{equation}\label{c.11}
\frac {{x,_r}^2 + {y,_r}^2} {{\cal E}^2} = \frac {{t,_r}^2} {\Phi^2} - \frac
{{\Phi_1}^2} {(\varepsilon - k) {\Phi^2}}.
\end{equation}

In order to write the equations in a compact way it is convenient to use the
symbol $\Psi$ introduced in (\ref{6.4}). Recall: this is a factor of shear, and
when it vanishes, the Szekeres model reduces to Friedmann. Thus, in searching
for RLPs in nontrivial models we will assume $\Psi \neq 0$. Using (\ref{c.11})
in (\ref{5.16}) adapted to RLPs we obtain
\begin{eqnarray}\label{c.12}
\tau,_r t,_r &=& \frac {\Phi_1 \Phi_{01} \tau} {\varepsilon - k} + \frac
{\Phi,_t {t,_r}^2 \tau} {\Phi} - \frac {\Phi,_t {\Phi_1}^2 \tau} {(\varepsilon -
k) \Phi} \nonumber \\
&\equiv& \frac {\Phi_1 \Psi \tau} {\varepsilon - k} + \frac {\Phi,_t {t,_r}^2
\tau} {\Phi}.
\end{eqnarray}
We also have:
\begin{eqnarray}
&& \Phi_{01} - \Phi,_t \Phi_1 / \Phi \equiv \Phi,_{tr} - \Phi,_t \Phi,_r / \Phi
=
\Psi, \label{c.13} \\
&& \Phi,_{ttr} - \Phi,_{tt} \Phi,_r / \Phi = \Psi,_t + \frac {\Phi,_t} {\Phi}
\Psi, \label{c.14} \\
&& \Phi,_{trr} - \Phi,_t \Phi,_{rr} / \Phi = \Psi,_r + \frac {\Phi,_r} {\Phi}
\Psi. \label{c.15}
\end{eqnarray}
Assuming $\Psi \neq 0$ we now multiply (\ref{6.2}) by $\Phi {\Phi_1}^2
t,_r/[(\varepsilon - k) \Psi]$, use (\ref{c.7}) -- (\ref{c.15}), cancel $\tau$
that factors out, and write the result in the form:
\begin{equation}\label{c.16}
x,_r \left({t,_r}^3 + c_2 {t,_r}^2 + c_3 t,_r + c_4\right) = B_1 {t,_r}^3 + B_3
t,_r,
\end{equation}
where
\begin{eqnarray}
c_2 &\df& \frac {2 \Phi \left(\Phi_1 \Psi,_t - \Psi^2\right)} {(\varepsilon - k)
\Psi}, \label{c.17} \\
c_3 &\df& \frac {- 3 {\Phi_1}^2 + \Phi \Phi_1 \Psi,_r / \Psi - \Phi \Phi,_{rr}}
{\varepsilon - k} \nonumber \\
&&\ \ \ \ \  + \frac {\Phi^2 {\cal E},_{rr} / {\cal E} + \Phi_1 \Phi,_r}
{\varepsilon - k}, \label{c.18} \\
c_4 &\df& \frac {2 \Phi {\Phi_1}^2 \Psi} {(\varepsilon - k)^2}, \label{c.19}
\\
B_1 &\df& \frac {P,_r y (2y - C)} {(\varepsilon - k) S^2}, \label{c.20} \\
B_3 &\df& - \frac {3 {\Phi_1}^2} {2 (\varepsilon - k)}\ B_1. \label{c.21}
\end{eqnarray}
Then, using (\ref{c.1}), (\ref{c.7}) and (\ref{c.11}) we can rewrite (\ref{4.9})
in the form:
\begin{equation}\label{c.22}
t,_{rr} = c_5 {t,_r}^3 + c_6 {t,_r}^2 + c_7 t,_r + c_8 + A x,_r t,_r,
\end{equation}
where
\begin{eqnarray}
c_5 &\df& - \frac {\varepsilon - k} {\Phi \Phi_1}, \label{c.23} \\
c_6 &\df& 2 \frac {\Psi} {\Phi_1} + \frac {\Phi,_t} {\Phi}, \label{c.24} \\
c_7 &\df& \frac {\Phi,_{rr} - \Phi {\cal E},_{rr} / {\cal E}} {\Phi_1} - \frac
{{\cal E},_r} {\cal E} + \frac {k,_r} {2 (\varepsilon - k)} + \frac {\Phi_1}
{\Phi}, \nonumber \\
 \label{c.25} \\
c_8 &\df& - \frac {\Psi \Phi_1} {\varepsilon - k}, \label{c.26} \\
A &\df& \frac {\left(4D_3 + C^ 2\right) P,_r y \Phi} {(2y - C) S^2 {\cal E}^2
\Phi_1}. \label{c.27}
\end{eqnarray}

Combining (\ref{c.9}) and (\ref{c.11}) we get:
\begin{equation}\label{c.28}
{x,_r}^2 = \frac {(2y - C)^2 {\cal E}^2} {\left(4D_3 + C^ 2\right) {\Phi^2}}
\left({t,_r}^2 - \frac {{\Phi_1}^2} {\varepsilon - k}\right).
\end{equation}
Equations (\ref{c.16}) and (\ref{c.28}) determine $\dril t r$ along the
hypothetic RLP. Formally, a solution for $\dril t r$ of these equations always
exists, but it must be consistent with the geodesic equations, and this is what
we will investigate next. Namely, every solution of these equations must be
preserved along the null geodesics. To see whether it is, we first transform
this set into a single polynomial equation for $\dril t r$.

We square (\ref{c.16}) and use (\ref{c.28}) in the result. We thus obtain an
8-th degree polynomial in $t,_r$, whose coefficient at ${t,_r}^8$ is
\begin{equation}\label{c.29}
\alpha_1 = \frac {(2y - C)^2 {\cal E}^2} {\left(4D_3 + C^ 2\right) {\Phi^2}}.
\end{equation}
It is seen that it cannot vanish except when $y = C/2$, but this defines a
``radial'' geodesic that exists only in the axially symmetric case
\cite{BKHC2009}. Thus we divide the 8-th degree polynomial by $\alpha_1$ and
obtain the following equation
\begin{eqnarray}\label{c.30}
&& {t,_r}^8 + 2 c_2 {t,_r}^7 + a_3 {t,_r}^6 + a_4 {t,_r}^5 + a_5 {t,_r}^4 + a_6
{t,_r}^3 \nonumber \\
&& \ \ \ \ \ \ + a_7 {t,_r}^2 + a_8 {t,_r} + a_9 = 0,
\end{eqnarray}
where:
\begin{eqnarray}
a_3 &\df& 2 c_3 + {c_2}^2 - \frac {{\Phi_1}^2} {\varepsilon - k} - \frac
{\left(4D_3 + C^ 2\right) \Phi^2 {P,_r}^2 y^2} {(\varepsilon - k)^2 S^4 {\cal
E}^2}, \nonumber \\
 \label{c.31} \\
a_4 &\df& 2 c_4 + 2 c_2 c_3 - 2 c_2 \frac {{\Phi_1}^2} {\varepsilon - k},
\label{c.32} \\
a_5 &\df& 2 c_2 c_4 + {c_3}^2 - \left(2 c_3 + {c_2}^2\right) \frac {{\Phi_1}^2}
{\varepsilon - k}  \nonumber \\
&& + 3 \frac {{\Phi_1}^2} {(\varepsilon - k)^3} \frac {\left(4D_3 + C^
2\right) \Phi^2 {P,_r}^2 y^2} {S^4 {\cal E}^2}, \label{c.33} \\
a_6 &\df& 2 c_3 c_4 - \left(2 c_4 + 2 c_2 c_3\right) \frac {{\Phi_1}^2}
{\varepsilon - k}, \label{c.34} \\
a_7 &\df& {c_4}^2 - \left(2 c_2 c_4 + {c_3}^2\right) \frac {{\Phi_1}^2}
{\varepsilon - k}  \nonumber \\
&& - \frac {9 {\Phi_1}^4} {4 (\varepsilon - k)^4} \frac {\left(4D_3 + C^
2\right) \Phi^2 {P,_r}^2 y^2} {S^4 {\cal E}^2}, \label{c.35} \\
a_8 &\df& - 2 c_3 c_4 \frac {{\Phi_1}^2} {\varepsilon - k}, \label{c.36} \\
a_9 &\df& - {c_4}^2 \frac {{\Phi_1}^2} {\varepsilon - k}. \label{c.37}
\end{eqnarray}
Now we differentiate (\ref{c.30}) along the null geodesic by the rule
(\ref{b.1}), and use (\ref{c.30}) to eliminate ${t,_r}^{10}$, ${t,_r}^9$ and
${t,_r}^8$ from the result. In this way we obtain:
\begin{eqnarray}\label{c.38}
&& b_1 {t,_r}^7 + b_2 {t,_r}^6 + b_3 {t,_r}^5 + b_4 {t,_r}^4 + b_5 {t,_r}^3 +
b_6 {t,_r}^2 \nonumber \\
&& \ \ \ \ \ \ + b_7 {t,_r} + b_8 \nonumber \\
&& + x,_r \left(\beta_1 {t,_r}^7 + \beta_2 {t,_r}^6 + \beta_3 {t,_r}^5 + \beta_4
{t,_r}^4 + \beta_5 {t,_r}^3\right. \nonumber \\
&& \ \ \ \ \ \ + \beta_6 {t,_r}^2 + \left.\beta_7 {t,_r} + \beta_8\right) = 0,
\end{eqnarray}
where
\begin{eqnarray}
b_1 &\df& 8 c_8 + 2 c_{2,r} + a_{3,t} - 2 a_3 c_6 - 3 a_4 c_5 - 2 c_2 c_7
\nonumber \\
&& - 4 c_2 c_{2,t} + 6 a_3 c_2 c_5 + 4 {c_2}^2 c_6 - 8 {c_2}^3 c_5,
\label{c.39} \\
b_2 &\df& a_{3,r} + a_{4,t} + 14 c_2 c_8 - 2 a_3 c_7 - 3 a_4 c_6 - 4 a_5 c_5
\nonumber \\
&& - 2 a_3 c_{2,t} + 2 a_4 c_2 c_5 + 2 a_3 c_2 c_6 + 2 {a_3}^2 c_5 \nonumber \\
&& - 4 a_3 {c_2}^2 c_5, \label{c.40} \\
b_3 &\df& a_{4,r} + a_{5,t} + 6 a_3 c_8 - 3 a_4 c_7 - 4 a_5 c_6 - 5 a_6 c_5
\nonumber \\
&& - 2 a_4 c_{2,t} + 2 a_5 c_2 c_5 + 2 a_4 c_2 c_6 + 2 a_3 a_4 c_5 \nonumber \\
&&  - 4 a_4 {c_2}^2 c_5, \label{c.41} \\
b_4 &\df& a_{5,r} + a_{6,t} + 5 a_4 c_8 - 4 a_5 c_7 - 5 a_6 c_6 - 6 a_7 c_5
\nonumber \\
&& - 2 a_5 c_{2,t} + 2 a_6 c_2 c_5 + 2 a_5 c_2 c_6 + 2 a_3 a_5 c_5 \nonumber \\
&&  - 4 a_5 {c_2}^2 c_5, \label{c.42} \\
b_5 &\df& a_{6,r} + a_{7,t} + 4 a_5 c_8 - 5 a_6 c_7 - 6 a_7 c_6 - 7 a_8 c_5
\nonumber \\
&& - 2 a_6 c_{2,t} + 2 a_7 c_2 c_5 + 2 a_6 c_2 c_6 + 2 a_3 a_6 c_5 \nonumber \\
&&  - 4 a_6 {c_2}^2 c_5, \label{c.43} \\
b_6 &\df& a_{7,r} + a_{8,t} + 3 a_6 c_8 - 6 a_7 c_7 - 7 a_8 c_6 - 8 a_9 c_5
\nonumber \\
&& - 2 a_7 c_{2,t} + 2 a_8 c_2 c_5 + 2 a_7 c_2 c_6 + 2 a_3 a_7 c_5 \nonumber \\
&&  - 4 a_7 {c_2}^2 c_5, \label{c.44} \\
b_7 &\df& a_{8,r} + a_{9,t} + 2 a_7 c_8 - 7 a_8 c_7 - 8 a_9 c_6 - 2 a_8 c_{2,t}
\nonumber \\
&& + 2 a_9 c_2 c_5 + 2 a_8 c_2 c_6 + 2 a_3 a_8 c_5 - 4 a_8 {c_2}^2 c_5,
\nonumber \\
 \label{c.45} \\
b_8 &\df& a_{9,r} + a_8 c_8 - 8 a_9 c_7 - 2 a_9 c_{2,t} + 2 a_9 c_2 c_6
\nonumber \\
&& + 2 a_3 a_9 c_5  - 4 a_9 {c_2}^2 c_5, \label{c.46} \\
\beta_1 &\df& 2 \left(c_{2,x} - \frac {2x} {2y - C}\ c_{2,y}\right) - 2 c_2 A,
\label{c.47} \\
\beta_2 &\df& a_{3,x} - \frac {2x} {2y - C}\ a_{3,y} - 2 a_3 A, \label{c.48} \\
\beta_3 &\df& a_{4,x} - \frac {2x} {2y - C}\ a_{4,y} - 3 a_4 A, \label{c.49} \\
\beta_4 &\df& a_{5,x} - \frac {2x} {2y - C}\ a_{5,y} - 4 a_5 A, \label{c.50} \\
\beta_5 &\df& a_{6,x} - \frac {2x} {2y - C}\ a_{6,y} - 5 a_6 A, \label{c.51} \\
\beta_6 &\df& a_{7,x} - \frac {2x} {2y - C}\ a_{7,y} - 6 a_7 A, \label{c.52} \\
\beta_7 &\df& a_{8,x} - \frac {2x} {2y - C}\ a_{8,y} - 7 a_8 A, \label{c.53} \\
\beta_8 &\df& a_{9,x} - \frac {2x} {2y - C}\ a_{9,y} - 8 a_9 A. \label{c.54}
\end{eqnarray}

We provisionally assume that the coefficient of $x,_r$ in (\ref{c.16}) is
nonzero. (We will later come back to this point and investigate what happens
when it is zero.) Then we determine $x,_r$ from (\ref{c.16}) and substitute the
result in (\ref{c.38}). After multiplying out to get a polynomial in $t,_r$, we
again use (\ref{c.30}) to eliminate ${t,_r}^{10}$ and ${t,_r}^9$ (but not
${t,_r}^8$). Then we assume that the coefficient of ${t,_r}^8$, denoted $d_1$,
is nonzero (we will check the case $d_1 = 0$ later), and divide the equation by
$d_1$. In this way we obtain:
\begin{eqnarray}\label{c.55}
&& {t,_r}^8 + \delta_2 {t,_r}^7 + \delta_3 {t,_r}^6 + \delta_4 {t,_r}^5
+ \delta_5 {t,_r}^4 + \delta_6 {t,_r}^3 \nonumber \\
&& \ \ \ \ \ \ + \delta_7 {t,_r}^2 + \delta_8 t,_r + \delta_9 = 0,
\end{eqnarray}
where $\delta_i \df d_i/d_1$, $i = 2, \dots, 9$, and
\begin{eqnarray}
d_1 &\df& b_3 - b_2 c_2 + b_1 c_3 + \beta_1 B_3  + \beta_3 B_1 - a_3 b_1
\nonumber \\
&& - a_3 \beta_1 B_1 - 2 \beta_2 B_1 c_2 + 2 b_1 {c_2}^2 + 4 \beta_1 B_1
{c_2}^2,\nonumber \\
&& \label{c.56} \\
d_2 &\df& b_4 + b_1 c_4 + b_3 c_2 + b_2 c_3 + \beta_2 B_3  + \beta_4 B_1 - a_4
b_1 \nonumber \\
&& - a_3 b_2 - a_4 \beta_1 B_1 - a_3 \beta_2 B_1 + a_3 b_1 c_2 \nonumber \\
&& + 2 a_3 \beta_1 B_1 c_2, \label{c.57} \\
d_3 &\df& b_5 + b_2 c_4 + b_4 c_2 + b_3 c_3 + \beta_3 B_3  + \beta_5 B_1 - a_5
b_1 \nonumber \\
&& - a_4 b_2 - a_5 \beta_1 B_1 - a_4 \beta_2 B_1 + a_4 b_1 c_2 \nonumber \\
&& + 2 a_4 \beta_1 B_1 c_2, \label{c.58} \\
d_4 &\df& b_6 + b_3 c_4 + b_5 c_2 + b_4 c_3 + \beta_4 B_3  + \beta_6 B_1 - a_6
b_1 \nonumber \\
&& - a_5 b_2 - a_6 \beta_1 B_1 - a_5 \beta_2 B_1 + a_5 b_1 c_2 \nonumber \\
&& + 2 a_5 \beta_1 B_1 c_2, \label{c.59} \\
d_5 &\df& b_7 + b_4 c_4 + b_6 c_2 + b_5 c_3 + \beta_5 B_3  + \beta_7 B_1 - a_7
b_1 \nonumber \\
&& - a_6 b_2 - a_7 \beta_1 B_1 - a_6 \beta_2 B_1 + a_6 b_1 c_2 \nonumber \\
&& + 2 a_6 \beta_1 B_1 c_2, \label{c.60} \\
d_6 &\df& b_8 + b_5 c_4 + b_7 c_2 + b_6 c_3 + \beta_6 B_3  + \beta_8 B_1 - a_8
b_1 \nonumber \\
&& - a_7 b_2 - a_8 \beta_1 B_1 - a_7 \beta_2 B_1 + a_7 b_1 c_2 \nonumber \\
&& + 2 a_7 \beta_1 B_1 c_2, \label{c.61} \\
d_7 &\df& b_6 c_4 + b_8 c_2 + b_7 c_3 + \beta_7 B_3 - a_9 b_1 - a_8 b_2
\nonumber \\
&& - a_9 \beta_1 B_1 - a_8 \beta_2 B_1 + a_8 b_1 c_2 + 2 a_8 \beta_1 B_1 c_2,
\nonumber \\
&& \label{c.62} \\
d_8 &\df& b_7 c_4 + b_8 c_3 + \beta_8 B_3 - a_9 b_2 - a_9 \beta_2 B_1 \nonumber
\\
&& + a_9 b_1 c_2 + 2 a_9 \beta_1 B_1 c_2, \label{c.63} \\
d_9 &\df& b_8 c_4. \label{c.64}
\end{eqnarray}

Every solution of (\ref{c.30}) is a candidate RLP, and every RLP must obey
(\ref{c.30}). Equation (\ref{c.55}) is the condition that (\ref{c.30}) is
preserved along null geodesics. Thus, every solution of (\ref{c.30}) must also
be a solution of (\ref{c.55}). Since (\ref{c.30}) and (\ref{c.55}) are of the
same degree in $t,_r$, it follows that both must have the same set of zeros.
Consequently, their coefficients must be the same. After we make sure that they
are the same, we may next investigate which zeros define RLPs. Thus, the
following equations are necessary conditions for the existence of RLPs:
\begin{eqnarray}\label{c.65}
&&2 c_2 = \delta_2, \qquad a_i = \delta_i, \quad i = 3, \dots 9, \nonumber \\
&& \Longleftrightarrow 2 c_2 d_1 - d_2 = 0, \qquad d_1 a_i - d_i = 0.
\end{eqnarray}
By far the simplest condition, as seen from (\ref{c.64}), is the one with $i =
9$. Even so, further calculations are so complicated and involve intermediate
equations so large that they could be done only using the computer algebra
system Ortocartan \cite{Kras2001, KrPe2000}, and we only describe how they were
done.

First we observe that the functions $(\Phi_1, y, {\cal E})$ are linearly
independent, and in the resulting final equation can be used as independent
variables. Although the proof is a simple exercise, it requires careful
inspection of special cases that we had earlier excluded for separate
investigation, so we give it in the separate Appendix \ref{simpex}.

The condition (\ref{c.65}) corresponding to $i = 9$ is
\begin{equation}\label{c.66}
d_1 a_9 - d_9 = 0.
\end{equation}
In this, one has to do the whole cascade of substitutions, listed in
(\ref{c.17}) -- (\ref{c.64}). In the result, we use (\ref{c.5}) -- (\ref{c.6}).
However, we use the last of (\ref{c.5}) only to eliminate ${\cal E},_{rr}$. For
${\cal E},_r$ we substitute from (\ref{4.1}), i.e.
\begin{equation}\label{c.67}
{\cal E},_r = {\cal E} \left(\Phi,_r - \Phi_1\right) / \Phi,
\end{equation}
in order to express ${\cal E},_r$ through $\Phi_1$.

We then use (\ref{c.15}) to express $\Phi,_{trr}$ through $\Psi$, and
(\ref{6.4}) to express $\Phi,_{tr}$ through $\Psi$. In the result we use
(\ref{c.2}) to eliminate $x^2$ and (\ref{c.4}) to express $x$ through ${\cal
E}$. From the resulting equation we can factor out ${\Phi_1}^6$, and we must
multiply it by ${\cal E}^4$ to get rid of negative powers of ${\cal E}$. The
final equation thus obtained has on its l.h.s. a polynomial of 4th degree in
${\cal E}$, of 4th degree in $y$ and of 6th degree in $\Phi_1$ (recall, we
determined that $(\Phi_1, y, {\cal E})$ are independent variables).\footnote{The
whole equation would take 1830 print lines on paper.} So, if this polynomial is
to be zero, the coefficients of all powers of the independent variables must
vanish separately.

Now we take this large polynomial as data for a second program, in which
$(\Phi_1, y, {\cal E})$ are treated as independent variables, no longer as
functions. In it, we determine the coefficient of $y^4$ and substitute ${\cal E}
= 0$ to find the term independent of ${\cal E}$. The resulting equation is:
\begin{equation}\label{c.68}
16 \left(4D_3 + C^2\right)^2 \Phi^6 \Psi \Psi,_t {P,_r}^4 / \left[S^8
(\varepsilon - k)^9\right] = 0.
\end{equation}
In this equation we can discard several alternatives: $\Phi = 0$ obviously,
$4D_3 + C^2 = 0$ because of (\ref{c.3}), $\Psi = 0$ because it defines the
Friedmann limit and $P,_r = 0$ because, in view of (\ref{b.11}) and the
paragraph above (\ref{b.22}), it leads to the $G_3/S_2$ symmetric cases
considered in Appendix \ref{RLPinG3S2}. The only case to consider is thus
$\Psi,_t = 0$.

In order to investigate it, we substitute $\Psi,_t = 0$ in the large main
polynomial, and in the resulting somewhat smaller polynomial we take the
coefficient of $y^4$. The equation that results is:
\begin{eqnarray}\label{c.69}
&& - 16 \left(4D_3 + C^2\right) \left(4P^2 + C^2\right) {\cal E} \Phi^5 \Psi^3
\nonumber \\
&&\ \ \ \ \ \ \  \times {P,_r}^3 / \left[P S^7 (\varepsilon - k)^9\right] = 0.
\end{eqnarray}
The factors ${\cal E}$ and $\Phi$ obviously cannot vanish, and why zero values
of $\left(4D_3 + C^2\right)$, $\Psi$ and $P,_r$ are discarded was explained
above. Thus, the only case left is $4P^2 + C^2 = 0$, which means $P = 0 = C$.
But this is just a special case of $P,_r = 0$ discarded above. Thus,
(\ref{c.68}) does not include any case that would define any new RLP, apart from
those considered elsewhere.

We go back to (\ref{c.55}) to consider the case $d_1 = 0$. The calculation is
almost the same as we did for (\ref{c.66}), with only minor differences: this
time the expression is somewhat simpler, and ${\Phi_1}^6$ does not factor out.
We employ the algebraic program to calculate ${\cal E}^4 d_1$, with the same
cascade of substitutions as before, take the coefficient of $y^4$ at ${\cal E} =
0$, and obtain an equation almost identical to (\ref{c.68}):
\begin{equation}\label{c.70}
4 \left(4D_3 + C^2\right)^2 \Phi^4 \Psi,_t {P,_r}^4 / \left[S^8 (\varepsilon -
k)^4 \Psi\right] = 0.
\end{equation}
As explained above, only $\Psi,_t$ could possibly be zero, so we substitute
$\Psi,_t = 0$ in the main large polynomial, and in the resulting expression take
the coefficient of $y^4$. The result is almost the same as (\ref{c.69})
\begin{eqnarray}\label{c.71}
&& - 4 \left(4D_3 + C^2\right) \left(4P^2 + C^2\right) {\cal E} \Phi^3 \Psi
\nonumber \\
&&\ \ \ \ \ \ \  \times {P,_r}^3 / \left[P S^7 (\varepsilon - k)^4\right] = 0,
\end{eqnarray}
and again does not include any case that would define a new RLP.

Finally, we go back to (\ref{c.54}), where we assumed that the coefficient of
$x,_r$ in (\ref{c.16}) was nonzero, and investigate what happens when it is
zero. Then
\begin{equation}\label{c.72}
B_1 {t,_r}^3 + B_3 t,_r = 0,
\end{equation}
and one of the solutions of this is $t,_r = 0$. This we immediately discard
because it defines a spacelike curve, while our RLPs must be null geodesics. In
consequence of (\ref{c.21}), another solution of (\ref{c.72}) is $B_1 = 0$. But
this implies $P,_r = 0$ or $y = 0$ or $y = C/2$. The first case leads to the
$G_3/S_2$ solutions considered in Appendix \ref{RLPinG3S2}, the other two to the
axially symmetric solutions of Appendix \ref{axialRLP}. So the only possibility
left to fulfil (\ref{c.72}) is
\begin{equation}\label{c.73}
t,_r = \pm \sqrt{\frac 3 {2 (\varepsilon - k)}}\ \Phi_1.
\end{equation}
Putting this into the coefficient of $x,_r$ in (\ref{c.16}) we get:
\begin{eqnarray}\label{c.74}
&& \left[\frac 3 {2 (\varepsilon - k)}\right]^{3/2} \Phi_1 \left\{\left[\mp 1 +
\frac {2^{3/2} \Phi \Psi,_t} {\sqrt{3 (\varepsilon - k)} \Psi}\right]
{\Phi_1}^2\right. \nonumber \\
&& \ \ \ \ + \left[- \frac {2^{3/2} \Phi \Psi} {3 \sqrt{3 (\varepsilon - k)}}
\pm \frac 2 3 \left(\Phi \frac {\Psi,_r} {\Psi} + \Phi,_r\right)\right] \Phi_1
\nonumber \\
&& \ \ \ \ \left.\pm \frac 2 3 \Phi \left(\Phi \frac {{\cal E},_{rr}} {\cal E}
-\Phi,_{rr}\right)\right\} = 0.
\end{eqnarray}
In this expression, we do the same series of substitutions that we did in the
large polynomial that resulted from (\ref{c.66}): we express ${\cal E},_{rr}$
through ${\cal E},_r$ using (\ref{c.5}), ${\cal E},_r$ through $\Phi,_r$ and
$\Phi_1$ using (\ref{c.67}), then $x$ through ${\cal E}$ and $y$ using
(\ref{c.4}), and multiply the whole expression by ${\cal E}$. The result is:
\begin{eqnarray}\label{c.75}
&& \left[\mp 1 + \frac {2^{3/2} \Phi \Psi,_t} {\sqrt{3 (\varepsilon - k)}
\Psi}\right] {\cal E} {\Phi_1}^2 \nonumber \\
&&+ \left[- \frac {2^{3/2} \Phi \Psi} {3 \sqrt{3 (\varepsilon - k)}} \pm \frac 2
3 \left(\Phi \frac {\Psi,_r} {\Psi} + \Phi,_r + \Phi \frac {S,_r}
S\right)\right] {\cal E} \Phi_1 \nonumber \\
&& \mp \frac 2 3 \Phi \Phi,_{rr} {\cal E} \mp \frac 2 3 \Phi^2 \left[\left(\frac
{P,_r} S\right),_r \left(- \frac {S {\cal E}} P + \frac {Cy} {2P} + \frac {D_3}
P\right)\right. \nonumber \\
&& \ \ \ \ \ \  + \left.\left(\frac {S,_r} S\right),_r {\cal E} + \frac {S,_r
\Phi,_r} {S \Phi} {\cal E}\right] = 0.
\end{eqnarray}
We then use the fact that $(\Phi_1, y, {\cal E})$ are linearly independent and
require that each coefficient of an independent function is zero. There is only
one term containing $y$, with the coefficient $C (P,_r / S),_r$. But $C$ cannot
be zero, as explained below (\ref{c.3}). Thus $(P,_r / S),_r = 0$ is the unique
implication of this (it will be seen from the following that we need not
consider whether this condition is consistent with the other equations that $P$
and $S$ must obey). This means:
\begin{equation}\label{c.76}
P,_r = \alpha_0 S,
\end{equation}
where $\alpha_0$ is an arbitrary constant. Taking this into account, and taking
the term independent of $\Phi_1$ in (\ref{c.75}) we get:
\begin{equation}\label{c.77}
\Phi,_{rr} + \Phi \left(\frac {S,_r} S\right),_r + \frac {S,_r \Phi,_r} S = 0.
\end{equation}
Using (\ref{c.76}), this is easily integrated with the result:
\begin{equation}\label{c.78}
\Phi = \frac {\chi_1(t) P} {\alpha_0 S} + \frac {\chi_2(t)} S,
\end{equation}
where $(\chi_1(t), \chi_2(t))$ are arbitrary functions of $t$. Both appear as
integration ``constants'' of (\ref{c.77}). Using such $\Phi$ in the definition
of $\Psi$, (\ref{6.4}), we get:
\begin{equation}\label{c.79}
\Psi \Phi = \frac {\gamma(t)} S, \qquad \gamma(t) \df \chi_2 \chi_{1,t} - \chi_1
\chi_{2,t}.
\end{equation}
But with $\Psi = \gamma(t) / (S \Phi)$ the last three terms in the coefficient
of $\Phi_1$ in (\ref{c.75}) sum up to zero, and what remains of that coefficient
is the equation $\Phi \Psi = 0$. The only solution of this can be $\Psi = 0$,
but we know it leads to the Friedmann model.

Consequently, the coefficient of $x,_r$ in (\ref{c.16}) is always nonzero.

Since (\ref{c.68}) and (\ref{c.70}) were, in their respective branches of the
calculation, among the necessary conditions for the existence of RLPs, we
conclude that the special quasi-hyperbolic Szekeres solution defined by
(\ref{c.1}) does not contain any RLPs except (possibly) when it becomes axially
symmetric or $G_3/S_2$, but these cases are considered in Appendices
\ref{axialRLP} and \ref{RLPinG3S2}.

\section{Proof that $(\Phi_1, y, {\cal E})$ are linearly
independent}\label{simpex}

\setcounter{equation} {0}

We take the equation
\begin{equation}\label{d.1}
\alpha \Phi_1 + \beta y + \gamma {\cal E} = 0
\end{equation}
with constant coefficients $\alpha, \beta, \gamma$ and prove that it implies
$\alpha = \beta = \gamma = 0$.

We substitute for $\Phi_1$ from (\ref{4.1}), then multiply the equation by $2 S
{\cal E}$ and use (\ref{c.2}) to eliminate $x^2$. We thus obtain a polynomial of
degree 1 in $x$ and degree 2 in $y$, which we denote by ${\cal P}$. We take the
second derivative of ${\cal P}$ by $xy$. The result is:
\begin{equation}\label{d.2}
2P (\beta S - \gamma C) = 0.
\end{equation}
We discard the solution $P = 0$ because this implies constant $S$ (see
(\ref{b.11}) and the remarks above (\ref{b.22})), and then the metric acquires a
$G_3/S_2$ symmetry group -- these are discussed in Appendix \ref{RLPinG3S2}. We
also discard the case $\beta \neq 0$ because then $S$, and consequently $P$, are
constant, again leading to the $G_3/S_2$ case. So finally, the implication of
(\ref{d.2}) is
\begin{equation}\label{d.3}
\beta = 0 = \gamma C.
\end{equation}

With this, we go back to ${\cal P}$ and take its second derivative by $y$. The
result is $-4 \gamma P^2 / S = 0$, and the solution of this is $\gamma = 0$. We
again go back to ${\cal P}$ with $\beta = \gamma = 0$, and take its first
derivative by $y$. The result is:
\begin{equation}\label{d.4}\alpha C \left(\Phi,_r + \Phi S,_r/S\right) = 0.
\end{equation}
When the expression in parentheses vanishes, $\Phi$ becomes a product of the
form $R(t)/ S(r)$, where $R(t)$ is an arbitrary function. Such form of $\Phi$
defines the Friedmann limit, in which we know that all null geodesics are RLPs,
so we discard this case. Thus, we follow the case $\alpha C = 0$.

Putting this, together with $\beta = \gamma = 0$, in ${\cal P}$ and taking the
derivative of the result by $x$ we obtain
\begin{equation}\label{d.5}
2 \alpha \left(- P \Phi,_r + P,_r \Phi - \Phi P S,_r / S\right) = 0.
\end{equation}
If the expression in parentheses should vanish, then the solution is $\Phi =
R(t) P(r) / S(r)$, which again leads back to the Friedmann model. Thus, finally,
(\ref{d.5}) implies $\alpha = 0$, which completes the proof.

\section{The RLPs with $P = 0$ and $x = 0$ along the geodesic.}\label{specB}

\setcounter{equation}{0}

We can leave aside the case when $Q,_r = 0$ because then the metric is axially
symmetric from the beginning.

It is useful to turn this case back to that of Appendix \ref{emptygen} by the
transformation (\ref{b.14}) -- (\ref{b.15}) with $a \neq 0 \neq b$. After the
transformation we have  $x' = - by'/a$, i.e. $\dril {x'} r \neq 0$ if $\dril
{y'} r \neq 0$, and $\widetilde{P} = - b \widetilde{Q} / a$, i.e.
$\widetilde{P},_r \neq 0$ if $\widetilde{Q},_r \neq 0$. Thus the new
$\widetilde{P}$ and $\widetilde{Q}$ obey (\ref{b.8}) with $D_0 = 0$ from the
beginning, and the RLP condition reduces to (\ref{b.11}) alone, with $(P, Q)$
replaced by $(\widetilde{P}, \widetilde{Q})$. The rest of the reasoning of
Appendix \ref{emptygen} then applies, unchanged, to $(\widetilde{P},
\widetilde{Q})$, with the same result:

\medskip

{\bf Corollary 3:}

\medskip

RLPs with $P = 0$ and $x = 0$ along the geodesic may exist only in the special
case when the coordinates may be transformed so that $Q = 0$ as well, i.e. the
metric is axially symmetric, or has a 3-dimensional symmetry group.

\section{The axially symmetric case $P = Q = 0$: only the axial geodesics $x,_r
= y,_r = 0$ are RLPs}\label{axialRLP}

\setcounter{equation}{0}

We know from Ref. \cite{BKHC2009} that in the quasi-spherical case $\varepsilon
= +1$ null geodesics on which $x$ and $y$ are constant exist only when the
Szekeres model is axially symmetric. Then coordinates may be chosen so that $P =
Q = 0$, and the constant-$(x, y)$ null geodesics have $x = y = 0$, i.e.
intersect each $t =$ constant space on the symmetry axis.

In this appendix we show that the statement above applies also with $\varepsilon
= 0$ and $\varepsilon = -1$, that the constant-$(x, y)$ null geodesics are RLPs,
and that other RLPs may exist only when the Szekeres spacetime has more
symmetries.

\subsection{Constant-$(x, y)$ null geodesics exist only in the axially symmetric
case}

This thesis was proven in Ref. \cite{BKHC2009} for $\varepsilon = +1$. The
assumption made there in the proof, that ${\cal E} > 0$, does not hold for
$\varepsilon = 0$ and $\varepsilon = -1$, so we first verify what happens when
${\cal E} = 0$.

It is seen from (\ref{4.10}) and (\ref{4.11}) that constant $(x, y)$ imply
$E_{12} = E_{13} = 0$ along the geodesic. With ${\cal E} = 0$ (\ref{4.4}) and
(\ref{4.5}) then imply that either ${\cal E},_r = 0$ at all $r$, which means a
$G_3/S_2$ symmetry (discussed in Appendix \ref{RLPinG3S2}), or ${\cal E},_x =
{\cal E},_y = 0$ along the geodesic, which means $P$ and $Q$ being constant,
i.e. axial symmetry. Thus, ${\cal E} = 0$ along a constant-$(x, y)$ null
geodesic implies axial symmetry anyway.\footnote{Moreover, as shown in Ref.
\cite{HeKr2008}, the location ${\cal E} = 0$ is infinitely far from any point
within the spacetime, i.e. does not in fact belong to the spacetime.}

The equations of Sec. 3.3.1 in Ref. \cite{BKHC2009} that are imposed on $(P, Q,
S)$ by the condition of constant $(x, y)$ along the geodesic become subcases of
our (\ref{b.8}) and (\ref{b.11}) for a general $\varepsilon$. As shown in our
Appendix \ref{emptygen}, they imply axial symmetry for any $\varepsilon$. This
is true even for the special solution discussed in Appendix \ref{specmetric}, as
we now show.

When $x,_r = y,_r = 0$ along a null geodesic, as stated above, (\ref{4.10}) and
(\ref{4.11}) imply $E_{12} = E_{13} = 0$ along this geodesic. Then (\ref{c.1})
implies that either $(i)$ $P,_r = 0$, or $(ii)$ $x = 0$ and $y^2 + D_3 = 0$, or
$(iii)$ $y = 0$ and $x^2 - D_3 = 0$. Case $(i)$ is axially symmetric. Case
$(ii)$ implies ${\cal E} = 0$ along the geodesic, and this was discussed above.
Case $(iii)$ implies $D_3 \geq 0$. However, the solution of Appendix
\ref{specmetric} has $D_3 \leq 0$ by definition. So the only subcase to consider
here is $D_3 = 0 \Longrightarrow x = 0$ along this geodesic. But then we have
again ${\cal E} = 0$, which completes the proof.

\subsection{Constant-$(x, y)$ null geodesics are RLPs}

As stated above, along null geodesics of constant $(x, y)$ we have $E_{12} =
E_{13} = 0$. Then (\ref{6.2}) and (\ref{6.3}) are fulfilled identically, which
means that such geodesics are RLPs.

\subsection{Other RLPs may exist in the axially symmetric case only with higher
symmetries}

We will now show that, in the axially symmetric case, (\ref{6.2}) --(\ref{6.3})
may have other solutions than constant $(x, y)$ only when the spacetime has more
symmetries than just the axial.

The whole reasoning and calculation is closely analogous to the one presented in
Appendix \ref{specmetric} for the special Szekeres solution. We proved in
Appendix \ref{emptygen} that in the axially symmetric case coordinates may be
chosen so that $P = Q = 0$ and the candidate RLP has $x = 0$. Then:
\begin{eqnarray}\label{f.1}
{\cal E} &=& \frac {x^2 + y^2} {2S} + \frac 1 2\ \varepsilon S, \nonumber \\
E_{12} &=& \varepsilon x S,_r / S, \qquad E_{13} =  \varepsilon y S,_r / S.
\end{eqnarray}
Note that with $\varepsilon = 0$, this axially symmetric Szekeres solution is in
fact plane symmetric. Thus, it will be considered together with other $G_3/S_2$
symmetric solutions in Appendix \ref{RLPinG3S2}. From here on in the present
appendix we assume $\varepsilon \neq 0$, i.e. $\varepsilon = \pm 1$.

With (\ref{f.1}) obeyed, (\ref{6.2}) and (\ref{6.5}) are fulfilled identically
along $x = 0$. From (\ref{5.15}) we obtain:
\begin{equation}\label{f.2}
\frac {{y,_r}^2} {{\cal E}^2} = \frac {{t,_r}^2} {\Phi^2} - \frac {{\Phi_1}^2}
{(\varepsilon - k) {\Phi^2}},
\end{equation}
and (\ref{c.12}) applies unchanged. We then multiply (\ref{6.3}) by $\Phi
{\Phi_1}^2 t,_r/[(\varepsilon - k) \Psi]$ and use (\ref{f.2}), (\ref{c.12}) --
(\ref{c.15}) and $x = 0$ in the result. As before, $\tau$ factors out and is
cancelled, and we obtain an equation almost identical to (\ref{c.16}), with the
same definitions of $(B_3, c_2, c_3, c_4)$, but with $y,_r$ in place of $x,_r$,
and with the definition of $B_1$ changed to:
\begin{equation}\label{f.3}
B_1 \df \frac {2 \varepsilon y S,_r} {(\varepsilon - k) S}.
\end{equation}

We proceed in strict analogy to Appendix \ref{emptygen}. From (\ref{4.9}) using
(\ref{f.2}) we again obtain (\ref{c.22}), with $y,_r$ in place of $x,_r$, and
with the same definitions of $(c_5, \dots, c_8)$, but with the definition of $A$
changed to:
\begin{equation}\label{f.4}
A \df \frac {2 \varepsilon y \Phi S,_r} {S {\cal E}^2 \Phi_1}.
\end{equation}
Then we square the current analogue of (\ref{c.16}) and use (\ref{f.2}) to
eliminate ${y,_r}^2$ from the result. We obtain an 8-th degree polynomial in
$t,_r$, but this time the coefficient of ${t,_r}^8$ is
\begin{equation}\label{f.5}
\alpha_1 = {\cal E}^2 / \Phi^2
\end{equation}
and is sure to be nonzero. Dividing the polynomial by $\alpha_1$ we obtain
(\ref{c.38}) again, but with the definitions of some of the coefficients changed
as follows:
\begin{eqnarray}
a_3 &\df& 2 c_3 + {c_2}^2 - \frac {{\Phi_1}^2} {\varepsilon - k} - \frac {4
\varepsilon^2 \Phi^2 {S,_r}^2 y^2} {(\varepsilon - k)^2 S^2 {\cal E}^2},
\label{f.6} \\
a_5 &\df& 2 c_2 c_4 + {c_3}^2 - \left(2 c_3 + {c_2}^2\right) \frac {{\Phi_1}^2}
{\varepsilon - k}  \nonumber \\
&& + \frac {12 \varepsilon^2 \Phi^2 {\Phi_1}^2 {S,_r}^2 y^2} {(\varepsilon -
k)^3 S^2 {\cal E}^2}, \label{f.7} \\
a_7 &\df& {c_4}^2 - \left(2 c_2 c_4 + {c_3}^2\right) \frac {{\Phi_1}^2}
{\varepsilon - k} - \frac {18 \varepsilon^2 \Phi^2 {\Phi_1}^4 {S,_r}^2 y^2}
{(\varepsilon - k)^4 S^2 {\cal E}^2}, \nonumber \\
&& \label{f.8}
\end{eqnarray}
the remaining ones are the same as given by (\ref{c.32}), (\ref{c.34}) and
(\ref{c.36}) -- (\ref{c.37}).

Now we differentiate the current analogue of (\ref{c.30}) by $r$ along the null
geodesic by the rule (\ref{b.1}). This time, however, $x = 0$ along our
candidate RLP, so no coefficient depends on $x$. We then use our analogue of
(\ref{c.30}) to eliminate ${t,_r}^{10}$, ${t,_r}^9$ and ${t,_r}^8$ from the
result. The equation that emerges is an analogue of (\ref{c.38}) with $y,_r$ in
place of $x,_r$, with the same definitions of $(b_1, \dots, b_8)$, and with the
definitions of $(\beta_1, \dots, \beta_8)$ changed to
\begin{eqnarray}
\beta_1 &\df& 2 c_{2,y} - 2 c_2 A, \label{f.9} \\
\beta_2 &\df& a_{3,y} - 2 a_3 A, \label{f.10} \\
\beta_3 &\df& a_{4,y} - 3 a_4 A, \label{f.11} \\
\beta_4 &\df& a_{5,y} - 4 a_5 A, \label{f.12} \\
\beta_5 &\df& a_{6,y} - 5 a_6 A, \label{f.13} \\
\beta_6 &\df& a_{7,y} - 6 a_7 A, \label{f.14} \\
\beta_7 &\df& a_{8,y} - 7 a_8 A, \label{f.15} \\
\beta_8 &\df& a_{9,y} - 8 a_9 A, \label{f.16}
\end{eqnarray}
where for $A$ the definition (\ref{f.4}) must be used.

Here we can assume that the coefficient of $y,_r$ in the present analogue of
(\ref{c.16}) is nonzero -- the explanation given in the paragraphs containing
(\ref{c.72}) -- (\ref{c.75}) still applies, except that the $B_1$ given by
(\ref{f.3}) cannot vanish for somewhat different reasons.\footnote{The cases
$\varepsilon = 0$ and $S,_r = 0$ define metrics of higher symmetry, treated in
Appendix \ref{RLPinG3S2}, while $y = 0$ (together with $x = 0$ assumed
throughout this appendix) defines an axial geodesic, which we already know is an
RLP.} Then we determine $y,_r$ from that equation and substitute the result in
the current analogue of (\ref{c.38}). After multiplying out to get a polynomial
in $t,_r$, we again use (\ref{c.30}) to eliminate ${t,_r}^{10}$ and ${t,_r}^9$
(but not ${t,_r}^8$). Then we assume that the coefficient of ${t,_r}^8$, denoted
$d_1$, is nonzero (we will check the case $d_1 = 0$ later), and divide the
equation by $d_1$. In this way we obtain an exact copy of (\ref{c.55}), with the
same definitions (\ref{c.56}) -- (\ref{c.64}) of the coefficients; but it is to
be remembered that some of the symbols in these formulae (namely $B_1$, $B_2$,
$a_3$, $a_5$, $a_7$ and all of $(\beta_1, \dots, \beta_8)$) now have different
definitions from those in Appendix \ref{specmetric}.

Consequently, eqs. (\ref{c.65}) must still hold, and we again choose
(\ref{c.66}) to investigate, by exactly the same method as before. By the method
of Appendix \ref{simpex} we show that $(\Phi_1, y, {\cal E})$ are still linearly
independent in the present case (i.e. with ${\cal E}$ given by (\ref{f.1}), and
along the $x = 0$ geodesic). The explanation given under (\ref{c.67}) still
applies, with the modification that now $x$ is nowhere present, so does not have
to be eliminated. In place of (\ref{c.68}) we now obtain:
\begin{equation}\label{f.17}
256 \varepsilon^4 \Phi^6 \Psi \Psi,_t {S,_r}^4 / \left[S^4 (\varepsilon -
k)^9\right] = 0.
\end{equation}
We recall that we excluded the case $\varepsilon = 0$ (since it is treated in
Appendix \ref{RLPinG3S2}), and $\Psi = 0$ because it reduces the Szekeres model
to Friedmann. We can also exclude $S,_r = 0$ because then the metric acquires a
$G_3/S_2$ symmetry and is also treated in Appendix \ref{RLPinG3S2}. So, as
before, the only case left to investigate is $\Psi,_t = 0$.

We substitute $\Psi,_t = 0$ in the large main polynomial, and in the resulting
smaller polynomial we take the coefficient of $y^4$. The equation that results
is:
\begin{equation}\label{f.18}
256 \varepsilon^3 {\cal E} \Phi^5 \Psi^3 {S,_r}^3 / \left[S^4 (\varepsilon -
k)^9\right] = 0.
\end{equation}
The only factors that could vanish here are $\Psi$ and $S,_r$, but, as explained
above, their vanishing leads to simpler cases of higher symmetry. Thus,
(\ref{f.18}) does not include any case that would define any new RLP, apart from
those considered elsewhere.

We go back to the paragraph after (\ref{f.16}) to consider the case $d_1 = 0$.
The explanation given above (\ref{c.70}) still applies, but this time, in the
expression ${\cal E}^4 d_1$ calculated by the algebraic program, we take the
coefficient of $y^4$ at ${\cal E} = 0$, and obtain:
\begin{equation}\label{f.19}
64 \varepsilon^4 \Phi^4 \Psi,_t {S,_r}^4 / \left[S^4 (\varepsilon - k)^4
\Psi\right] = 0.
\end{equation}
As explained above, only $\Psi,_t$ could possibly be zero, so we substitute
$\Psi,_t = 0$ in the main large polynomial, and in the resulting expression take
the coefficient of $y^4$. The result is:
\begin{equation}\label{f.20}
64 \varepsilon^3 {\cal E}^5 \Phi^3 \Psi {S,_r}^3 / \left[S^4 (\varepsilon -
k)^4\right] = 0,
\end{equation}
and again does not include any case that would define a new RLP.

So, the final conclusion is:

\medskip

{\bf Corollary 4:}

\medskip

In the axially symmetric Szekeres solutions, apart from cases of higher
symmetry, the only RLPs are the axial null geodesics that intersect each 3-space
of constant $t$ on the symmetry axis.

\section{There are no non-radial RLPs in any $G_3/S_2$ model.}\label{RLPinG3S2}

\setcounter{equation}{0}

We will investigate the equation $\chi = 0$ (see (\ref{7.1})) and will show that
it has no solutions defining nonradial RLPs, unless the model reduces to
Friedmann.

Several equations in this Appendix follow from the corresponding ones in the
Appendix \ref{specmetric} as the special case ${\cal E},_r = 0$; they are
similar but not identical.

We will use all equations adapted to the special case discussed in Sec.
\ref{LTredshift}, i.e. $\zeta = \psi = \xi = \eta = {\cal E},_r = \dril y r =
0$. From (\ref{5.15}) we find
\begin{equation}\label{g.1}
\frac {{x,_r}^2} {{\cal E}^2} = \frac {{t,_r}^2} {\Phi^2} - \frac {{\Phi_r}^2}
{(\varepsilon - k) \Phi^2}.
\end{equation}
Then, using (\ref{g.1}) in (\ref{5.16}) we obtain
\begin{equation}\label{g.2}
\tau,_r t,_r = \frac {\tau \Phi_r \Psi} {\varepsilon - k} + \frac {\tau \Phi,_t}
{\Phi} {t,_r}^2,
\end{equation}
where $\Psi$ is defined by (\ref{6.4}) -- as seen from (\ref{2.9}) this is a
coefficient of shear, whose vanishing defines the Friedmann limit.

We now substitute (\ref{g.1}) and (\ref{g.2}) in $\chi = 0$, where $\chi$ is
given by (\ref{7.1}). We multiply the result by $\Phi {\Phi,_r}^2 t,_r /
[(\varepsilon - k) \Psi]$, and cancel $\tau$ that factors out. The result is:
\begin{equation}\label{g.3}
W_1 \df {t,_r}^3 + c_2 {t,_r}^2 + c_3 t,_r + c_4 = 0,
\end{equation}
where
\begin{eqnarray}
c_2 &\df& \frac {2 \Phi \left(\Phi,_r \Psi,_t - \Psi^2\right)} {(\varepsilon -
k) \Psi} \label{g.4} \\
c_3 &\df& \frac {\Phi \left(\Phi,_r \Psi,_r - \Phi,_{rr} \Psi\right) - 2
{\Phi,_r}^2 \Psi} {(\varepsilon - k) \Psi} \label{g.5} \\
c_4 &\df& \frac {2 \Phi {\Phi,_r}^2 \Psi} {(\varepsilon - k)^2}. \label{g.6}
\end{eqnarray}

Adapting (\ref{4.9}) to the $G_3/S_2$ case, eliminating $x,_r$ with use of
(\ref{g.1}) and using (\ref{g.3}) to eliminate ${t,_r}^3$ we obtain:
\begin{equation}\label{g.7}
t_{rr} = c_6 {t,_r}^2 + c_7 t,_r + c_8,
\end{equation}
where
\begin{eqnarray}
c_6 &\df& \frac {2 \Psi,_t} {\Psi} + \frac {\Phi,_t} {\Phi}, \label{g.8} \\
c_7 &\df& \frac {\Psi,_r} {\Psi} - \frac {\Phi,_r} {\Phi} + \frac {k,_r} {2
(\varepsilon - k)}, \label{g.9} \\
c_8 &\df& \frac {\Phi,_r \Psi}  {\varepsilon - k}. \label{g.10}
\end{eqnarray}
Now we differentiate (\ref{g.3}) along a null geodesic (since the equation must
hold all along it), by the rule given in (\ref{b.1}), and use (\ref{g.7}) to
eliminate $t,_{rr}$. The resulting equation is of 4-th degree in $t,_r$. We
eliminate the 4th power of $t,_r$ by using (\ref{g.3}). In the end, we obtain an
equation of degree 3 in $t,_r$, which we write symbolically as follows:
\begin{equation}\label{g.11}
d_1 {t,_r}^3 + d_2 {t,_r}^2 + d_3 t,_r + d_4 = 0.
\end{equation}
The expressions for the coefficients in (\ref{g.11}) are:
\begin{eqnarray}
&&d_1 = c_{2,t} + 3 c_7 - c_2 c_6, \label{g.12} \\
&&d_2 = c_{2,r} + c_{3,t} + 3 c_8 + 2 c_2 c_7 - 2 c_3 c_6, \label{g.13} \\
&&d_3 = c_{3,r} + c_{4,t} + 2 c_2 c_8 + c_3 c_7 - 3 c_4 c_6, \label{g.14} \\
&&d_4 = c_{4,r} + c_3 c_8. \label{g.15}
\end{eqnarray}

For the beginning, let us assume that $d_1 \neq 0$. For further considerations
it will be more convenient to write (\ref{g.11}) as follows:
\begin{equation}\label{g.16}
W_2 \df {t,_r}^3 + \delta_2 {t,_r}^2 + \delta_3 t,_r + \delta_4 = 0,
\end{equation}
where $\delta_i \df d_i/d_1$, $i = 2, 3, 4$.

Equation (\ref{g.3}) is equivalent to $\chi = 0$ in (\ref{7.1}). Thus
(\ref{g.3}), just like (\ref{7.1}), defines the collection of RLPs together with
the conditions of their existence. Every solution of (\ref{g.3}) and (\ref{7.1})
is a candidate RLP, and every RLP must obey (\ref{g.3}) and (\ref{7.1}). Then,
(\ref{g.16}) is the condition that the solutions of (\ref{g.3}) are preserved
along the null geodesics, thus every solution of (\ref{g.3}) must be a solution
of (\ref{g.16}) and vice versa. But if (\ref{g.3}) and (\ref{g.16}) have the
same set of solutions, then they must be identical, i.e. their respective
coefficients must be equal. Thus, the necessary conditions for the existence of
RLPs are:
\begin{equation}\label{g.17}
c_i = \delta_i \Longleftrightarrow c_i d_1 - d_i = 0, \quad i = 2, 3, 4.
\end{equation}

As with the previous calculations, we employed the algebraic program Ortocartan
\cite{Kras2001, KrPe2000}. We consider (\ref{g.17}) with $i = 4$, the simplest
one. In it, we substitute (\ref{g.4}) -- (\ref{g.15}) and multiply the result by
$(\Psi / \Phi,_r)$ to get a polynomial in $\Phi$, $\Psi$ and their derivatives
($\Phi,_r$ factors out in the original expression). The resulting expression is
simple enough to be shown here:
\begin{eqnarray}\label{g.18}
&& W_3 = \frac {- \Phi \Phi,_r \Psi^2 k,_r + 8 \Phi^2 \Phi,_r \Psi^2 \Psi,_t +
4 \Phi \Phi,_t {\Phi,_r}^2 \Psi \Psi,_t}
{(\varepsilon - k)^3} \nonumber \\
&& \ \ \ \ \ + \frac {4 \Phi^2 {\Phi,_r}^2 \Psi \Psi,_{tt} - 12 \Phi^2
{\Phi,_r}^2 {\Psi,_t}^2} {(\varepsilon - k)^3} \nonumber \\
&& \ \ \ \ \ + \frac {3 \Phi \Phi,_r \Psi \Psi,_r - 6 {\Phi,_r}^2 \Psi^2 - 3
\Phi \Phi,_{rr} \Psi^2} {(\varepsilon - k)^2} = 0.
\end{eqnarray}
We assume $\Lambda = 0$ and take the $k > 0$ model for the beginning. The
solution of (\ref{2.3}) can then be written as
\begin{eqnarray}\label{g.19}
\Phi (t,r) &=& \frac M k (1 - \cos \eta), \nonumber \\
\eta - \sin \eta &=& \frac {k^{3/2}} M \left[t - t_B(r)\right],
\end{eqnarray}
where $\eta$ is a parameter (dependent on $t$ and $r$), and $t_B(r)$ is an
arbitrary function, the bang time. We introduce the abbreviations:
\begin{equation}\label{g.20}
I_M \df \frac {3 k,_r} {2k} - \frac {M,_r} M, \qquad D_M \df \frac {k^2 t_{B,r}}
M.
\end{equation}
The derivatives of $\Phi$ and $\Psi$ can then be written as
\begin{eqnarray}
&& \Phi,_r = \left(\frac M k\right),_r (1 - \cos \eta) + \frac {M I_M} k  \frac
{\sin \eta (\eta - \sin \eta)} {1 - \cos \eta} \nonumber \\
&&\ \ \ \ \ \ \ \  - \frac {M D_M} {k^{3/2}} \frac {\sin \eta} {1 - \cos \eta},
\label{g.21} \\
&& \Phi,_t = \frac {\sqrt{k} \sin \eta} {1 - \cos \eta},\label{g.22} \\
&& \Phi_{rr} = \left(\frac M k\right),_{rr} (1 - \cos \eta) \nonumber \\
&&\ \ \ \ \ + \left(\frac M k\right),_r \sin \eta \left[2 I_M \frac {\eta - \sin
\eta} {1 - \cos \eta} - \frac {D_M} {\sqrt{k} (1 - \cos \eta)}\right] \nonumber
\\
&&\ \ \ \ \ + \frac {M (I_M),_r} k \frac {\sin \eta (\eta - \sin \eta)} {1 -
\cos \eta} \nonumber \\
&&\ \ \ \ \ + \frac {M {I_M}^2} k \frac {(2 \sin \eta - \sin \eta \cos \eta -
\eta) (\eta - \sin \eta)} {(1 - \cos \eta)^2} \nonumber \\
&&\ \ \ \ \ - \frac {M I_M D_M} {k^{3/2}} \frac {3 \sin \eta - \sin \eta \cos
\eta - 2 \eta} {(1 - \cos \eta)^2} \nonumber \\
&&\ \ \ \ \ - \left(\sqrt{k} t_{B,r}\right),_r \frac {\sin \eta} {1 - \cos \eta}
- \frac {M {D_M}^2} {k^2 (1 - \cos \eta)^2}, \label{g.23} \\
&& \Psi = \sqrt{k} I_M \frac {3 \sin \eta - \eta \cos \eta - 2 \eta} {(1 - \cos
\eta)^2} + D_M \frac {2 + \cos \eta} {(1 - \cos \eta)^2}, \nonumber \\
\label{g.24} \\
&& \Psi,_r = \left(\sqrt{k} I_M\right),_r \frac {3 \sin \eta - \eta \cos \eta -
2 \eta} {(1 - \cos \eta)^2} \nonumber \\
&&\ \ \ \ \ + I_M \frac {4 \cos \eta + \eta \sin \eta (5 + \cos \eta) - 4 - 4
\sin^2 \eta} {(1 - \cos \eta)^4} \nonumber \\
&&\ \ \ \ \ \ \ \ \ \ \ \ \ \ \ \times \left[\sqrt{k} I_M (\eta - \sin \eta) -
D_M\right] \nonumber \\
&&\ \ \ \ \ + \left(D_M\right),_r \frac {2 + \cos \eta} {(1 - \cos \eta)^2
\vspace{4mm}} \nonumber \\
\nonumber \\
&&\ \ \ \ \ - D_M \frac {\sin \eta (5 + \cos \eta)} {(1 - \cos \eta)^4}
\left[I_M (\eta - \sin \eta) - \frac {D_M} {\sqrt{k}}\right], \nonumber \\
\label{g.25} \\
&& \Psi,_t = \frac {k^2 I_M} M \frac {4 \cos \eta + \eta \sin \eta (5 + \cos
\eta) - 4 - 4 \sin^2 \eta} {(1 - \cos \eta)^4} \nonumber \\
&&\ \ \ \ \ - \frac {k^{3/2} D_M} M \frac {\sin \eta (5 + \cos \eta)} {(1 - \cos
\eta)^4}, \label{g.26} \\
&& \Psi,_{tt} = \frac {k^{7/2} I_M} {M^2 (1 - \cos \eta)^5}\ \left(-23 \eta - 19
\eta \cos \eta \right. \nonumber \\
&& \ \ \ \ \ \ \ \ \ \ \ \left.+ 2 \eta \sin^2 \eta + 33 \sin \eta + 9 \sin \eta
\cos \eta\right) \nonumber \\
&& \ \ \ - \frac {k^3 D_M} {M^2} \times \frac {-19 \cos \eta + 2 \sin^2 \eta -
23} {(1 - \cos \eta)^5}. \label{g.27}
\end{eqnarray}

{}From here on, the intermediate expressions become so large that we cannot
reproduce them here; we only describe how the calculation is done. We substitute
(\ref{g.21}) -- (\ref{g.27}) in (\ref{g.18}) and multiply the result by $(1 -
\cos \eta)^6$ to get a polynomial in $(1 - \cos \eta)$ (of 6th degree). It is
also a polynomial of 4th degree in $\eta$. The function $\eta$ is the only one
in this polynomial that depends on $t$, the coefficients of $\eta$, $(1 - \cos
\eta)$ and their powers depend only on $r$. Thus we treat $\eta$ and $(1 - \cos
\eta)$ as independent variables, and the coefficients of their different powers
must vanish separately.

In $W_3$ transformed in this way we now take the coefficient of $\eta^4$, the
resulting equation is a polynomial of degree 4 in $(1 - \cos \eta)$:
\begin{eqnarray}\label{g.28}
&& \frac {M^2 {I_M}^4} {(\varepsilon - k)^3} \left[9 (\varepsilon/k - 1) (1 -
\cos \eta)^4\right. \nonumber \\
&&\ \ \ \ \ \ \ \ \ \ - 78 (\varepsilon/k - 1) (1 - \cos \eta)^3 \nonumber \\
&&\ \ \ \ \ \ \ \ \ \ + (216 \varepsilon/k - 276) (1 - \cos \eta)^2  \nonumber \\
&&\ \ \ \ \ \ \ \ \ \ \left.- (189 \varepsilon/k - 381) (1 - \cos \eta) -
144\right] = 0.\ \ \ \ \ \ \ \ \ \
\end{eqnarray}
The term independent of $(1 - \cos \eta)$ clearly shows that ${I_M} = 0$ is a
unique solution of this, independently of the value of $k$.

So, we substitute ${I_M} = 0$ in the main large polynomial and in the resulting
smaller polynomial we take the term independent of $(1 - \cos \eta)$. The
equation that results is:
\begin{equation}\label{g.29}
- 144 M^2 {D_M}^4 / [k^2 (\varepsilon - k)^3] = 0.
\end{equation}
Here the unique solution is $D_M = 0$. But with $I_M = D_M = 0$ we get $\Psi =
0$ from (\ref{g.24}), i.e. the Friedmann limit. Thus, $c_4 d_1 - d_4 = 0$, which
is one of the necessary conditions for the existence of RLPs, can in this case
be fulfilled only when the Szekeres model trivializes to Friedmann.

The calculation above was done for $k > 0$. The calculation with $k < 0$ is
essentially the same and need not be done separately -- it is enough to replace
$(k, \eta)$ in all equations with $(- \widetilde{k}, {\rm i} \widetilde{\eta})$.

When $k = 0$, we have necessarily $\varepsilon = +1$ and the calculation must be
done separately. Then we have:
\begin{eqnarray}\label{g.30}
&& \Phi = \left(\frac {9 M} 2\right)^{1/3} \left(t - t_B\right)^{2/3},
\nonumber \\
&& \Psi = \frac 2 3 \left(\frac {9 M} 2\right)^{1/3} t_{B,r} \left(t -
t_B\right)^{-4/3}.
\end{eqnarray}
With the $r$-coordinate chosen so that $M = M_0 r^3$, where $M_0$ is a constant,
this simplifies $W_3$ in (\ref{g.18}) to
\begin{eqnarray}\label{g.31}
&& W_4 \df - (64/9) {M_0}^2 r^6 {t_{B,r}}^4 \left(t - t_B\right)^{-4} \nonumber
\\
&&\ \ \ \ \ - (14/3) \times 36^{1/3} {M_0}^{4/3} r^4 {t_{B,r}}^4 \left(t -
t_B\right)^{-10/3} \nonumber \\
&&\ \ \ \ \ + (256/3) {M_0}^2 r^5 {t_{B,r}}^3 \left(t - t_B\right)^{-3}
\nonumber \\
&&\ \ \ \ \ + 14 \times 36^{1/3} {M_0}^{4/3} r^3
{t_{B,r}}^3 \left(t - t_B\right)^{-7/3} \nonumber \\
&&\ \ \ \ \ \parbox{1mm}{\vspace{9mm}} - 112 {M_0}^2 r^4 {t_{B,r}}^2 \left(t -
t_B\right)^{-2} \\
&&\ \ \ \ \ - 3 \times 36^{1/3} {M_0}^{4/3} r^2 {t_{B,r}}^2 \left(t -
t_B\right)^{-4/3} \nonumber \\
&&\ \ \ \ \ + 3 \times 36^{1/3} {M_0}^{4/3} r^3 t_{B,r r}t_{B,r} \left(t -
t_B\right)^{-4/3} = 0.  \nonumber
\end{eqnarray}

Now the independent variables are $t$ and $r$, and $t$ appears always in the
combination $(t - t_B)$. Thus different powers of $(t - t_B)$ are linearly
independent, and their coefficients must vanish separately. Whichever term we
take, except for the last two, the result is always the same:
\begin{equation}\label{g.32}
t_{B,r} = 0
\end{equation}
(because $M_0 = 0$ is the vacuum, i.e. Schwarzschild, limit of the L--T model).
This guarantees that all the terms in (\ref{g.31}) vanish. However, as seen from
(\ref{g.30}), $t_{B,r} = 0$ means $\Psi = 0$, i.e. zero shear (see (\ref{6.4})
and (\ref{2.9})), i.e. the Friedmann limit.
Thus, there are no non-radial RLPs also when $k = 0$, which completes the proof
in the case $d_1 \neq 0$.

We go back to (\ref{g.15}), where we assumed $d_1 \neq 0$ and proceed from there
on to consider the case $d_1 = 0$. Instead of (\ref{g.18}) we now get:
\begin{eqnarray}\label{g.33}
&& W_5 = \frac {- 6 \Phi \Phi,_r {\Psi,_t}^2 / \Psi^2 + 2
\Phi \Phi,_r \Psi,_{tt} / \Psi} {\varepsilon - k} \nonumber \\
&&\ \ \ \ \  + \frac {4 \Phi \Psi,_t + 2 \Phi,_t \Phi,_r \Psi,_t / \Psi + (3/2)
k,_r} {\varepsilon - k} \nonumber \\
&&\ \ \ \ \ - 3 \frac {\Phi,_r} {\Phi} + 3 \frac {\Psi,_r} {\Psi} = 0.
\end{eqnarray}
As before, we begin by considering the case $k > 0$. We multiply $W_5$ by
$\Psi^2 \Phi (1 - \cos \eta)^6$ and substitute for $\Phi$ and $\Psi$ from
(\ref{g.19}) -- (\ref{g.27}). What results is a polynomial of degree 6 in $(1 -
\cos \eta)$ and of degree 3 in $\eta$. Taking the coefficient of $\eta^3$ we
obtain:
\begin{eqnarray}\label{g.34}
&& W_6 = \frac {M {I_M}^3 \sin \eta} {\varepsilon - k} \left[- 6 (\varepsilon -
k) (1 - \cos \eta)^3\right. \nonumber \\
&&\ \ \ \ \ \ + 45 (\varepsilon - k) (1 - \cos \eta)^2 - 81 (\varepsilon - k) (1
- \cos \eta) \nonumber \\
&&\ \ \ \ \ \left. + 30 k (1 - \cos \eta) - 36 k\right] = 0.
\end{eqnarray}
Looking at the term independent of $(1 - \cos \eta)$ we see that the unique
solution of this is $I_M = 0$.

So we substitute $I_M = 0$ in the main polynomial, and in the resulting
expression we take the term independent of $(1 - \cos \eta)$. The resulting
equation is:
\begin{equation}\label{g.35}
\frac {36 \sin \eta M {D_M}^3} {\sqrt {k} (\varepsilon - k)} = 0.
\end{equation}
The unique solution of this is $D_M = 0$, which, together with $I_M = 0$, leads
back to the Friedmann limit. Thus, no RLPs exist in nontrivial Szekeres
spacetimes in this case, either.

The argument given before, that the result for $k < 0$ follows by the
substitution $(k, \eta) \to (- \widetilde{k}, {\rm i} \widetilde{\eta})$, is
still valid. So we now consider $d_1 = 0$ with $k = 0$. We substitute $k = 0$
(and, as is necessary, $\varepsilon = +1$) in (\ref{g.33}), multiply the result
by $\Psi$, substitute then for $\Psi$ and $\Phi$ from (\ref{g.30}), and obtain:
\begin{eqnarray}\label{g.36}
&& W_7 = (16/9) M_0 r^3 {t_{B, r}}^2 \left(t - t_B\right)^{-3} \nonumber \\
&&\ \ \ \ \  + 2 \times 36^{1/3} {M_0}^{1/3} r {t_{B, r}}^2 \left(t -
t_B\right)^{-7/3} \nonumber \\
&&\ \ \ \ \ - (56/3) M_0 r^2 t_{B, r} \left(t - t_B\right)^{-2} \nonumber \\
&&\ \ \ \ \ + 36^{1/3} {M_0}^{1/3} r t_{B, r r} \left(t - t_B\right)^{-4/3} = 0.
\end{eqnarray}
The coefficients of independent powers of $\left(t - t_B\right)$ have to vanish
separately, as explained before. Whichever term we take, except for the last
one, the result is always the same:
\begin{equation}\label{g.37}
t_{B,r} = 0
\end{equation}
and this implies the Friedmann limit in the same way as explained after
(\ref{g.32}). This also guarantees that the whole of (\ref{g.36}) is fulfilled.

Thus, in every case considered, the assumption that non-radial RLPs could exist
leads to either the Friedmann limit or the Schwarzschild limit. The final
conclusion is that the only RLPs in the $G_3/S_2$ models are the radial null
geodesics. $\square$

\section{A detailed description of the model presented in Sec.
\ref{numex}.}\label{ModAlg}

\setcounter{equation}{0}

The algorithm used in the calculations discussed in Sec. \ref{numex} consists of
following steps:

\begin{enumerate}
\item First we set the observer at $R_0$ (the present-day areal distance)
and consider sources which are, at the present instant, away from the observer
by 1 Gly ($\approx 306.6$ Mpc).

\item To calculate the evolution of the model one needs to follow the following
points:

\begin{itemize}
 \item
The radial coordinate is chosen to be the areal radius at the present instant:
$\bar r = \Phi(t_{0},r)$. However, to simplify the notation we will omit the bar
and denote the new radial coordinate by $r$.

\item
The chosen asymptotic cosmic background is an open Friedman model, i.e.
$\Omega_m = 0.3$ and $\Lambda=0$. The background density is then given by

 \begin{equation}
\rho_b = \Omega_m \times \rho_{cr} = 0.3 \times \frac{3H_0^2}{8 \pi G} (1+z)^3,
 \end{equation}
where the Hubble constant is $H_0 =72$ km s$^{-1}$ Mpc$^{-1}$.

\item
The initial time $t_0$ is calculated from the following formula for the
background Friedman Universe

\begin{equation}
t(z) =  \int\limits_{0}^{\infty} \frac{ H_0^{-1} (1+\widetilde{z})^{-1}{\rm d}
\widetilde{z}}{\sqrt{ \Omega_{m} (1+\widetilde{z})^3 +
(1-\Omega_m)(1+\widetilde{z})^2}},
 \end{equation}

\item The age of the universe is assumed to be everywhere the same: $t_B =0$.

\item The function $M(r)$ follows from (\ref{2.5}), where the present-day
density is
\[ \rho(t_0,r) =  \rho_0 \left[ 1 + \delta - \delta \exp \left( -
{r^2}/{\sigma^2} \right) \right]\]

\item Because of the assumed spherical symmetry ${\rm e}^{\nu} = 1$.

\item The function $k(r)$ can be calculated from (\ref{2.6}).

\item Then the evolution of the model can be calculated from eq. (\ref{2.3}).
\end{itemize}

\item We then find a null geodesic that joins the observer and the source. The
angle between the direction towards the source and the direction towards the
origin, at the present instant, is denoted as $\gamma$.

\item The null geodesics are found in the following manner:
\begin{itemize}
\item Because of spherical symmetry we may set one of the angular components of
the null vector to zero. We set $k^\phi=0$.

\item The second angular component, $k^\theta$, follows from $R^2 k^\theta = J =
{\rm const} = R_0 \sin \gamma$, where $R_0 = R(t_0,r_0)$, i.e. at the observer's
position. This relation is a consequence of (\ref{4.10}) and (\ref{4.11}) and
was derived in \cite{BKHC2009} (see equation (3.26) in \cite{BKHC2009}).

\item The radial component is evaluated from (\ref{4.8})
with $E_{12} = 0 = E_{13} = {\cal E},_r$, and
$\Sigma = {\cal E}^2 ({\rm d} \theta/{\rm d} r)^2$.

\item The time component of the null vector is found from $k^\alpha k_\alpha =
0$.
\end{itemize}

\item We then find two other null geodesics: one that will reach the
observer in 1 Gy time, the other one that arrived at the observer's position 1
Gy ago. Because of the non-RLP effect these geodesics approach the observer at
angles that are different from $\gamma$.

\item The difference between these angles allows us to evaluate the rate of
change of the angle $\gamma$.
\end{enumerate}

{\bf Acknowledgments:} The work of AK was partly supported by the Polish
Ministry of Higher Education grant N N202 104 838. K.B. acknowledges the support
of the Marie Curie Fellowship under the grant PIEF-GA-2009-252950.


\bigskip

\begin{thebibliography}{99}

\bibitem{Bole2006} K. Bolejko, {\it Phys. Rev}. {\bf D73}, 123508 (2006).

\bibitem{Bole2007} K. Bolejko, {\it Phys. Rev}. {\bf D75}, 043508 (2007).

\bibitem{Bole2009a} K. Bolejko, {\it Gen. Rel. Grav.} {\bf 41}, 1585 (2009).

\bibitem{Bole2009b} K. Bolejko, {\it Gen. Rel. Grav.} {\bf 41}, 1737 (2009).

\bibitem{IRGW2008} M. Ishak, J. Richardson, D. Garred, D. Whittington, A.
Nwankwo, R. Sussman, {\it Phys. Rev.} {\bf D78}, 123531 (2008).

\bibitem{BC2010}
K. Bolejko and M.-N. C\'el\'erier, {\it Phys. Rev.} {\bf D82}, 103510 (2010).

\bibitem{NTI2010}
K. Bolejko and R. A. Sussman, Phys. Lett. B (2011) in press,
doi:10.1016/j.physletb.2011.02.007.

\bibitem{Lema1933} G. Lema\^{\i}tre, {\it Ann. Soc. Sci. Bruxelles} {\bf
A53}, 51 (1933); English translation, with historical comments (p. 637): {\it
Gen. Rel. Grav.} {\bf 29}, 641 (1997).

\bibitem{Tolm1934} R.C. Tolman, {\it Proc. Nat. Acad. Sci. USA} {\bf
20}, 169 (1934); reprinted, with historical comments (p. 931): {\it Gen. Rel.
Grav.} {\bf 29}, 935 (1997).

\bibitem{Bond1947} H. Bondi, {\it Mon. Not. Roy. Astr. Soc.} {\bf 107}, 410
(1947); reprinted, with historical comments (p. 1777), in {\it Gen. Rel. Grav.}
{\bf 31}, 1783 (1999).

\bibitem{PlKr2006} J. Pleba\'nski and A. Krasi\'nski, {\it An Introduction to
General Relativity and Cosmology}, Cambridge U P (2006).

\bibitem{BKHC2009} K. Bolejko, A. Krasi\'nski, C. Hellaby, M. N. C\'el\'erier,
{\it Structures in the Universe by exact methods -- formation, evolution,
interactions.} Cambridge University Press 2009.

\bibitem{NoDe2007} B. C. Nolan, U. Debnath, {\it Phys. Rev.} {\bf D76}, 104046
(2007).

\bibitem{Szek1975a} P. Szekeres, {\it Commun. Math. Phys}. {\bf 41}, 55 (1975).

\bibitem{Szek1975b} P. Szekeres, {\it Phys. Rev.} {\bf D12}, 2941 (1975).

\bibitem{KaSa1966} R. Kantowski and R. K. Sachs, {\it J. Math. Phys.}
{\bf 7}, 443 (1966).

\bibitem{BaSS1984} J. D. Barrow and J. Stein-Schabes, {\it Phys. Lett.} {\bf
A103}, 315 (1984).

\bibitem{GoWa1982} S. W. Goode and J. Wainwright, {\it Phys. Rev.} {\bf D26},
3315 (1982).

\bibitem{Kras1997} A. Krasi\'nski, {\it Inhomogeneous Cosmological
Models}, Cambridge U P (1997).

\bibitem{HeKr2008} C. Hellaby and A. Krasi\'nski, {\it Phys. Rev.} {\bf
D77}, 023529 (2008).

\bibitem{Kras2008} A. Krasi\'nski, {\it Phys. Rev.} {\bf D78}, 064038 (2008).

\bibitem{HeKr2002} C. Hellaby and A. Krasi\'nski, {\it Phys. Rev.} {\bf D66},
084011 (2002).

\bibitem{Hell1996} C. Hellaby, {\it Class. Quant. Grav.} {\bf 13}, 2537 (1996).

\bibitem{KS66}
J. Kristian and  R.K. Sachs, {\it Astroph. J.} {\bf 143}, 379 (1966); reprinted,
with historical comments, in {\it Gen. Relativ. Gravit.} {\bf 43}, 331 (2011).

\bibitem{BoWy2008}
K. Bolejko and J. S. B. Wyithe  {\it J. Cosmol.
Astropart. Phys}. {\bf 02(2009)}, 020 (2009).

\bibitem{Kras2001} A. Krasi\'nski, {\it Gen. Rel. Grav.} {\bf 33}, 145 (2001).

\bibitem{KrPe2000} A. Krasi\'nski, M. Perkowski, {\it The system ORTOCARTAN --
user's manual}. Fifth edition, Warsaw 2000.
\end{thebibliography}
\end{document}